\documentclass[12pt, draftclsnofoot, onecolumn]{IEEEtran}
\usepackage{graphicx}
\usepackage{epstopdf}
\usepackage{amsthm}
\usepackage{epsfig}
\usepackage{latexsym}
\usepackage{amsfonts}
\usepackage{here}
\usepackage{rawfonts}
\usepackage[latin1]{inputenc}
\usepackage[T1]{fontenc}
\usepackage{calc}
\usepackage{capitalgreekitalic}
\usepackage{url}
\usepackage{enumerate}
\usepackage{color}
\usepackage{url}
\usepackage[tbtags]{amsmath}
\usepackage{amssymb}
\usepackage{upref}
\usepackage{epic,eepic}
\usepackage{times}
\usepackage{dsfont}
\usepackage{comment}
\usepackage{cite}
\usepackage{bbm}
\usepackage{bm}

\newtheorem{theorem}{\bf Theorem}
\newtheorem{proposition}{\bf Proposition}
\newtheorem{lemma}{\bf Lemma}

\newtheorem{definition}{\bf Definition}

\usepackage{dsfont}

\newcounter{step}
\newlength{\totlinewidth}
\newenvironment{algorithm}{%
  \rule{\linewidth}{1pt}
  \begin{list}{}%
    {\usecounter{step}%
      \settowidth{\labelwidth}{\textbf{Step 2:}}%
      \setlength{\leftmargin}{\labelwidth}%
      \setlength{\topsep}{-2pt}%
      \addtolength{\leftmargin}{\labelsep}%
      \addtolength{\leftmargin}{2mm}%
      \setlength{\rightmargin}{2mm}%
      \setlength{\totlinewidth}{\linewidth}%
      \addtolength{\totlinewidth}{\leftmargin}%
      \addtolength{\totlinewidth}{\rightmargin}%
      \setlength{\parsep}{0mm}%
      \raggedright}}%
  {\end{list}%
  \rule{\linewidth}{1pt}}
\newcounter{substep}

  {\end{list}}

\newlength{\aligntop}
\newlength{\alignbot}
 \makeatletter
\renewenvironment{align}{%
  \vspace{\aligntop}
  \start@align\@ne\st@rredfalse\m@ne
}{%
  \math@cr \black@\totwidth@
  \egroup
  \ifingather@
    \restorealignstate@
    \egroup
    \nonumber
    \ifnum0=`{\fi\iffalse}\fi
  \else
    $$%
  \fi
  \ignorespacesafterend%
  \vspace{\alignbot}\par\noindent
} \makeatother

\IEEEoverridecommandlockouts

\usepackage{algorithm}
\usepackage{algorithmic}
\usepackage{array}
\usepackage{makeidx}
\makeindex

\makeatletter
\newcommand\semihuge{\@setfontsize\semihuge{19.3}{25}}
\makeatother

\makeatletter
\newcommand\semismall{\@setfontsize\semihuge{12.4}{15}}
\makeatother

\allowdisplaybreaks[4]
\usepackage{subfigure}

\begin{document}
\title{\LARGE Federated Learning for Task and Resource Allocation in Wireless High Altitude Balloon Networks}
\vspace{-0.2cm}

\author{{Sihua Wang,} \emph{Student Member, IEEE}, {Mingzhe Chen,} \emph{Member, IEEE} \\
\vspace{-0.2cm}
{Changchuan Yin}, \emph{Senior Member, IEEE}, Walid Saad, \emph{Fellow, IEEE}\\
\vspace{-0.2cm}
{Choong Seon Hong}, \emph{Senior Member, IEEE}, {Shuguang Cui}, \emph{Fellow}, \emph{IEEE},\\
\vspace{-0.2cm}
and {H. Vincent Poor}, \emph{Fellow}, \emph{IEEE}\\
\thanks{S. Wang and C. Yin are with the Beijing Laboratory of Advanced Information Network, and the Beijing Key Laboratory of Network System Architecture and Convergence, Beijing University of Posts and Telecommunications, Beijing 100876, China. Email: \protect\url{sihuawang@bupt.edu.cn;} ccyin@ieee.org.}
\thanks{M. Chen is with the Department of Electrical Engineering, Princeton University, Princeton, NJ, 08544, USA, and also with the Shenzhen Research Institute of Big Data (SRIBD) and the Future Network of Intelligence Institute (FNii), Chinese University of Hong Kong, Shenzhen, 518172, China, Email: \protect\url{mingzhec@princeton.edu}.}
\thanks{W. Saad is with the Wireless@VT, Bradley Department of Electrical and Computer Engineering, Virginia Tech, Blacksburg, VA, 24060, USA, Email: \protect\url{walids@vt.edu}.}
\thanks{C. S. Hong is with the Department of Computer Science and Engineering,
Kyung Hee University, South Korea, Email: \protect\url{cshong@khu.ac.kr}.}
\thanks{S. Cui is with the Shenzhen Research Institute of Big Data (SRIBD) and the Future Network of Intelligence Institute (FNii), Chinese University of Hong Kong, Shenzhen, 518172, China, Email: \protect\url{robert.cui@gmail.com}.}
\thanks{H. V. Poor is with the Department of Electrical Engineering, Princeton University, Princeton, NJ, 08544, USA, Email: \protect\url{poor@princeton.edu}.}}
\maketitle
\vspace{-1.5cm}
\begin{abstract}In this paper, the problem of minimizing energy and time consumption for task computation and transmission is studied in a mobile edge computing (MEC)-enabled balloon network. In the considered network, each user needs to process a computational task in each time instant, where high-altitude balloons (HABs), acting as flying wireless base stations, can use their powerful computational abilities to process the tasks offloaded from their associated users. Since the data size of each user's computational task varies over time, the HABs must dynamically adjust the user association, service sequence, and task partition scheme to meet the users' needs. This problem is posed as an optimization problem whose goal is to minimize the energy and time consumption for task computing and transmission by adjusting the user association, service sequence, and task allocation scheme. To solve this problem, a support vector machine (SVM)-based federated learning (FL) algorithm is proposed to determine the user association proactively. The proposed SVM-based FL method enables each HAB to cooperatively build an SVM model that can determine all user associations without any transmissions of either user historical associations or computational tasks to other HABs. Given the prediction of the optimal user association, the service sequence and task allocation of each user can be optimized so as to minimize the weighted sum of the energy and time consumption. Simulations with real data of city cellular traffic from the OMNILab at Shanghai Jiao Tong University show that the proposed algorithm can reduce the weighted sum of the energy and time consumption of all users by up to 16.1\% compared to a conventional centralized method.


\end{abstract}

{\renewcommand{\thefootnote}{\fnsymbol{footnote}}
\footnotetext{Some in the results of this paper are found in \cite{WSH2020ICC}, which has been accepted for presentation at IEEE ICC 2020.}}

\section{Introduction}
High altitude balloons (HABs) are attracting increasing attention for future wireless communication networks, owing to their low deployment expense and large coverage range \cite{XPM}. In particular, HABs can be used for various services including broadband Internet access, digital video/audio request, and emergency response \cite{ZCYX}. To provide these services to ground users, HABs, acting as relays between ground users and base stations (BSs) as done in \cite{MWMY,YZY,YQR,PGMM}, must transmit computational tasks that are generated by the ground users to terrestrial BSs or the cloud via wireless backhaul links. Since the wireless resources that can be used for relaying ground user data to far-away BSs is limited, it is impractical for HABs to transmit all of their computational tasks to the BSs or the cloud. In addition, long-haul transmissions will incur significant delays \cite{CKH}. To reduce the task transmission delay and enable the HABs to process computational tasks locally, one can deploy mobile edge computing (MEC) locally at each HAB \cite{YDP}. In particular, MEC-enabled HABs can directly process the computational tasks offloaded from the ground users without the need to transmit them to far-away BSs \cite{WMM}. However, deploying MEC at HABs also faces many challenges such as energy efficiency of processing computational tasks, user association, and computational task allocation.

A number of existing works have studied important problems related to task offloading and computational resource optimization such as in \cite{SOJ,YCK,MAWM,MAH}. In \cite{SOJ}, the authors studied the use of MEC-enabled unmanned aerial vehicles (UAVs) to service ground users. The authors in \cite{YCK} minimized the sum transmit power of UAVs and user via jointly optimizing users' transmit power and task allocation in an MEC network. In \cite{MAWM}, the authors derived the minimum number of UAVs that can cover a given space. The authors in \cite{MAH} studied the deployment of UAVs so as to maximize the number of service users. Compared to UAVs with limited flight time \cite{SOJ,YCK,MAWM,MAH}, HABs can be tethered and equipped with powerful computing resources and, hence, they can continuously hover to serve ground users. Meanwhile, HABs can be deployed in the stratosphere to reduce the energy cost for hovering \cite{XPMX}. Moreover, the existing works in \cite{SOJ,YCK,MAWM,MAH} do not consider scenarios in which the data size of computational tasks requested by each user changes over time. As the data size of each computational task varies, each HAB must dynamically adjust user association, service sequence, and task allocation to minimize the ground users' energy and time consumption. For this purpose, each HAB must collect the users' computational task information. However, each computational task processed by HABs is offloaded from a ground user and, hence, each HAB must first determine user association so as to collect the users' computational task information and adjust service sequence as well as task allocation. In addition, each HAB can only collect the information related to the computational tasks of its associated users instead of the computational information from all users. Therefore, given only the computational task information of a subset of users, each HAB must use traditional iterative or distributed optimization methods, such as Lagrangian dual decomposition \cite{HSCK} or game theory \cite{SHY}, to find the globally optimal user association, thus resulting in additional overhead and delay for computational task processing. Moreover, if such known techniques are used, as the data size of each computational task varies, the HABs must rerun their iterative or distribution optimization algorithm to cope with this change thus increasing the time needed to minimize the energy and time consumption of ground users. To tackle this challenge, each HAB needs to predict the user association based on the historical information of the computational tasks. One promising solution is to use machine learning algorithms for the prediction of optimal user association. In particular, ML algorithms can train a learning model to find a relationship between the future optimal user association and the computational task that each user is currently executing. Based on the predicted optimal user association, the HABs can optimize service sequence and task allocation hence minimizing the energy and time consumption of each user.

Recently, a number of existing works such as in \cite{RCWYB,GYZYJL,LNHDT} used machine learning algorithms to solve resource optimization problems related to MEC. The work in \cite{RCWYB} developed a deep learning method to optimize the user association scheme. In \cite{GYZYJL}, the authors sought to minimize the task processing delay using deep reinforcement learning. The authors in \cite{LNHDT} developed a cache and communication resource allocation scheme using a deep recurrent neural network. However, most of these works \cite{RCWYB,GYZYJL,LNHDT} use centralized learning algorithms. Hence, each distributed node needs to transmit its local dataset such as its historical association scheme and the data size of the requested task to a central controller for training a machine learning model. However, it is impractical to send all local datasets to a central controller in MEC-enabled HAB networks since the transmission of local datasets can lead to significant energy consumption. To address this challenge, one can use federated learning (FL) \cite{MUWYM} that enables distributed devices to collaboratively train a machine learning model via sharing trained parameters with other devices instead of massive dataset. In \cite{SSM}, the problem of joint transmission power and resource allocation is solved by FL to reduce the queuing delay of all users. The work in \cite{FLWC} introduced an energy-efficient strategy for transmission and computation resource allocation under delay constraints. In \cite{TRICC}, the authors proposed an FL algorithm to optimize resource allocation scheme in mobile edge computing. However, the existing works in \cite{SSM,FLWC,TRICC} that directly averaged the learning models generated by HABs in the FL training process did not optimize the parameters that are needed to capture the relationship between different learning models of different users, thus degrading the FL performance. Therefore, it is necessary to develop a novel FL algorithm that can capture the relationships among HABs' user association schemes and can be implemented over HABs.

The main contribution of this paper is a novel framework for dynamically optimizing the energy and time consumption of wireless users in an MEC-enabled HAB network based on accurate predictions of the user association. Our key contributions include:


\begin{itemize}
\item We consider an MEC-enabled HAB network, in which the users request computational tasks that can be of different data size over time. To provide computing services to their users, the HABs must dynamically determine the optimal user association, service sequence, and task allocation. This joint user association, service sequence and task allocation problem is formulated as an optimization problem whose goal is to minimize the weighted sum of the energy and time consumption of all users.


\item  To solve this optimization problem, an SVM-based FL algorithm is proposed to determine the user association proactively. The proposed SVM-based FL algorithm allows the HABs to cooperatively train an optimal SVM model that can predict the optimal user association without any transmissions of historical user association results nor of the data size of the task requested by each user. Given the predicted user association, the optimization problem of service sequence and task allocation can be simplified and solved.

\item  We perform fundamental analysis on the convergence of the proposed SVM-based FL algorithm, and we show that this algorithm converges to the optimal SVM model after training process. Meanwhile, our results also show that the learning rate and the target accuracy will significantly affect the convergence speed. By appropriately setting the learning rate and the target accuracy, the convergence speed of the proposed algorithm can be guaranteed.

\end{itemize}

Simulations using real data from the OMNILab at Shanghai Jiao Tong University show that the proposed algorithm can reduce the weighted sum of the energy and time consumption of all users by up to 16.1\% compared to the conventional centralized SVM method. To the best of our knowledge, this is the first work that \emph{studies the use of support vector machine (SVM)-based FL to dynamically determine user association so as to minimize the weighted sum of the energy and time consumption in an MEC-enabled HAB network}.

The rest of this paper is organized as follows. The system model and the problem formulation are described in Section II. Then, Section III discusses the proposed learning framework to predict user association. The optimization of service sequence and task allocation are determined in Section VI. Section V studied the convergence of the proposed algorithm. In Section VI, numerical results are presented and discussed. Finally, conclusions are drawn in Section VII.

\section{System Model and Problem Formulation}\label{se:system}
Consider an MEC-enabled HAB network that consists of a set $\mathcal{N}$ of $N$ HABs serving a set $\mathcal{M}$ of $M$ users over both uplink and downlink in a given geographical area. In this model, the users are associated with the HABs via wireless links and each HAB is equipped with computational resources to provide communication and computing services to the users. For example, HABs can be equipped with computational resources for analyzing the optimal route from the current location to the destination of each ground vehicle so as to provide navigation service to ground vehicles \cite{SWJW}. In this network, the uplink is used to transmit the computational task that each user offloads to the HAB while the downlink is used to transmit the computing result of the offloading task. We assume that the size of each task that user $m$ needs to process in each time instant $t$ is $z_{m,t}$, which will be changed as time elapses.
\begin{figure}[t]
\centering
\setlength{\belowcaptionskip}{-0.45cm}
\vspace{-0.1cm}
\includegraphics[width=10cm]{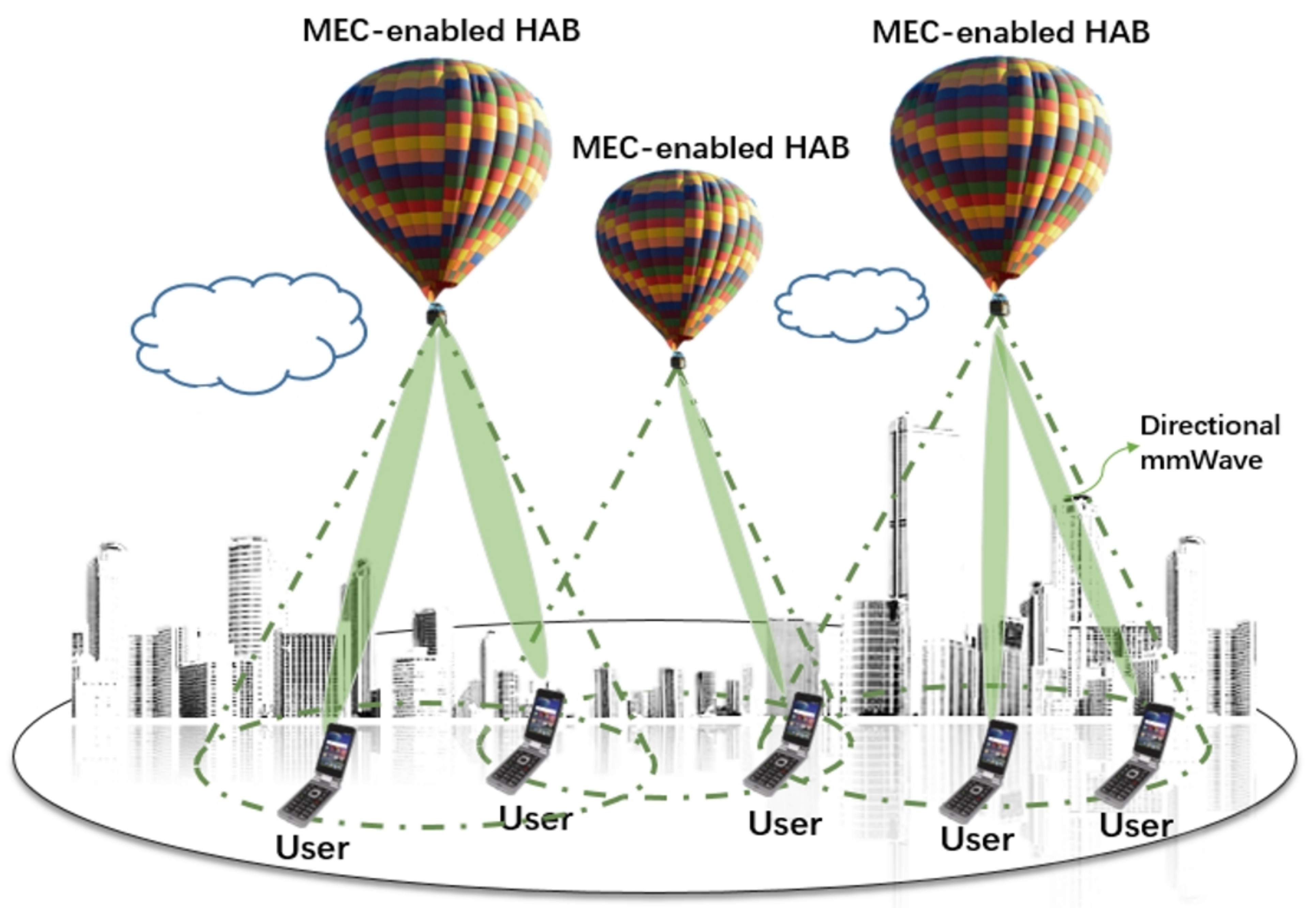}
\caption{An illustration of the considered MEC-enabled HAB network model.}
\vspace{-0.7cm}
\label{fig2}
\end{figure}

\subsection{Transmission Model}
In the considered scenario, all the communication links will use the millimeter wave (mmWave) frequency bands to provide high data rate services for ground users so as to satisfy the delay requirement of computational tasks \cite{OWM}. A time division multiple access (TDMA) scheme is adopted to support directional transmissions over the mmWave band. Note that, the channel gains of the mmWave links depend on the instantaneous large scale and small scale fading. For HAB-ground user transmission links (air-to-ground transmission links),  the large scale fading is the free space path loss and attenuation due to rain and clouds \cite{DD}. Small scale fading is modeled as Ricean fading due to the presence of line-of-sight rays from the HAB to most of the locations in the HAB service area \cite{EMMF}. The channel gains $g_{mn,t}$ and $h_{mn,t}$ between HAB $n$ and user $m$ over uplink and downlink during each time instant $t$ are given by:
\begin{equation}\label{eq:uplinkchannelgain}
\vspace{-0.05cm}
g_{mn,t}=\left( \frac{C}{4\pi r_{mn}f_c} \right)\!\cdot\! G_H(\Psi_{mn})\!\cdot\! G_m \!\cdot\! A(d_{mn})\!\cdot\! \varphi_{n,t},
\vspace{-0.05cm}
\end{equation}
\begin{equation}\label{eq:downlinkchannelgain}
\vspace{-0.05cm}
h_{mn,t}=\left( \frac{C}{4\pi r_{mn}f_c} \right)\!\cdot\! G_H(\Psi_{mn})\!\cdot\! G_m \!\cdot\! A(d_{mn})\!\cdot\! \varphi_{m,t},
\vspace{-0.05cm}
\end{equation}
respectively, where $C$ is the speed of light, $f_c$ is the carrier frequency, and $r_{mn}$ is the distance between HAB $n$ and user $m$; $G_H(\Psi_{mn})\!=\!\cos(\Psi_{mn})^{\rho}\frac{32{\rm log}2}{2(2\arccos(\sqrt[\rho]{0.5}))^2}$ is the gain seen at an angle $\Psi_{mn}$ between user $m$ and HAB $n$'s boresight axis with ${\rho}$ being the roll-off factor of the antenna. $G_m$ is the antenna gain of user $m$. $A(r_{mn})=10^{\left( \frac{3\chi r_{mn}}{10H}\right)}$ is the attenuation due to clouds and rain with $H$ being the HAB height and $\chi$ being the attenuation through the cloud and rain in dB/km. $\varphi_{n,t}$ and $\varphi_{m,t}$ represent the small scale Ricean gain during time instant $t$ for HAB $n$ and user $m$, respectively. Since a directional antenna is adopted at each HAB, the connectivity between HAB and user can be available for data transmission only if the directional antenna is directed towards each user and hence, interference is negligible. Given a bandwidth $B$ for each HAB, the rates of data transmission for uplink and downlink between user $m$ and HAB $n$ during time instant $t$ will be:
\begin{equation}\label{eq:uplinkdatarate}
\vspace{-0.1cm}
u_{mn,t}\left({a_{mn,t}}\right)=a_{mn,t} B{\log _2}\left(\!1\!+\! {\frac{{{P_\textrm{U}}{ g_{mn,t}}}}{{ \sigma^2 }}} \!\right),
\end{equation}
\begin{equation}\label{eq:downlinkdatarate}
\vspace{-0.1cm}
d_{mn,t}\left({a_{mn,t}}\right)=a_{mn,t} B{\log _2}\left(\!1\!+\! {\frac{{{P_\textrm{B}}{ h_{mn,t}}}}{{ \sigma^2 }}} \!\right),
\end{equation}
respectively, where $a_{mn,t}$ is the index of the user association with $a_{mn,t}=1$ indicating that user $m$ connects to HAB $n$ at time instant $t$, otherwise, we have $a_{mn,t}=0$. $P_\textrm{B}$ and $P_\textrm{U}$ are the transmit power of each HAB and user, which are assumed to be equal for all HABs and users, respectively. $\sigma^2$ represents the variance of the additive white Gaussian noise. The uplink and downlink transmission delay between user $m$ and HAB $n$ at time instant $t$ can be given by:
\vspace{0.1cm}
\begin{align}\label{eq:uplinkdelay}
l_{mn,t}^\textrm{U}\left( \beta_{mn,t},{a_{mn,t}}\right) =\frac{\beta_{mn,t} z_{m,t}}{u_{mn,t}\left({a_{mn,t}}\right)}, \quad
l_{mn,t}^\textrm{D}\left(\beta_{mn,t},{a_{mn,t}}\right)=\frac{\beta_{mn,t} z_{m,t}}{d_{mn,t}\left({a_{mn,t}}\right)},
\end{align}
respectively, where $\beta_{mn,t} z_{m,t}$ is the fraction of the task that user $m$ transmits to HAB $n$ for processing in each time instant $t$ with $\beta_{mn,t} \in [0,1]$ being the task division parameter.

\subsection{Computing Model}
In the considered model, each user's computational task can be cooperatively processed on the HAB, a process that we call \emph{edge computing}, or it can use \emph{local computing} on the user itself. Next, we introduce the models of edge computing and local computing in detail.


\subsubsection{Edge computing model}
Given the data size $\beta_{mn,t} z_{m,t}$ of the task that is offloaded from user $m$, the time used for HAB $n$ to process the task can be given by:
\vspace{-0.1cm}
\begin{equation}\label{eq:computingHAB}
l_{mn,t}^\textrm{B}\left( \beta_{mn,t}\right) =\frac{\omega^\textrm{B} \beta_{mn,t} z_{m,t}}{f^\textrm{B}},
\vspace{-0.1cm}
\end{equation}
where $f^\textrm{B}$ is the frequency of the central processing unit (CPU) clock of each HAB $n$ assumed to be equal for all HABs. $\omega^\textrm{B}$ is the number of CPU cycles required for computing data (per bit).
\subsubsection{Local computing model}
Given the data size $(1-\beta_{mn,t}) z_{m,t}$ of the task that is computed locally, the time used for user $m$ to process the task can be given by:
\vspace{-0.1cm}
\begin{equation}\label{eq:localcomputing}
l_{mn,t}^\textrm{L}\left( \beta_{mn,t}\right) =\frac{ \omega^\textrm{U}_m \left(1-\beta_{mn,t}\right) z_{m,t}}{f^\textrm{U}_m},
\vspace{-0.1cm}
\end{equation}
where $f^\textrm{U}_m$ is the frequency of the CPU clock of user $m$ and $\omega^\textrm{U}_m$ is the number of CPU cycles required for computing the data (per bit) of user $m$.

\subsection{Time Consumption Model}
In the proposed model, since users and HABs can process their computational task simultaneously, the total time used for the task computation is determined by the maximum time between the local computing time and edge computing time. Thus, based on (\ref{eq:uplinkdelay})--(\ref{eq:localcomputing}), the time needed by user $m$ and HAB $n$ to cooperatively process the computational task of user $m$ can be given by:
\vspace{-0.1cm}
\begin{equation}\label{eq:totaltime}
l_{mn,t}\!\left( \beta_{mn,t},\!a_{mn,t}\right)\!\!=\!\max\!\left\{ l_{mn,t}^\textrm{U}\!\left( \beta_{mn,t},\!{a_{mn,t}}\right)\!+\!l_{mn,t}^\textrm{B}\left( \beta_{mn,t}\right)\!\!+\!l_{mn,t}^\textrm{D}\left( \beta_{mn,t},\!{a_{mn,t}}\right)\!, l_{mn,t}^\textrm{L}\left( \beta_{mn,t}\right)\!\right\}\!,
\vspace{-0.1cm}
\end{equation}
where $l_{mn,t}^\textrm{U}\left( \beta_{mn,t},a_{mn,t}\right)\!+\!l_{mn,t}^\textrm{B}\left( \beta_{mn,t}\right)\!+\!l_{mn,t}^\textrm{D}\left( \beta_{mn,t},a_{mn,t}\right)$ represents the edge computing time and $l_{mn,t}^\textrm{L}\left( \beta_{mn,t}\right)$ represents the local computing time.

Moveover, since TDMA is used in the considered model, each user must wait for service, thus incurring a wireless access delay. For a given user $m$ that is associated with HAB $n$, this access delay can be given by:
\vspace{-0.2cm}
\begin{equation}\label{eq:queue}
l^{\rm S}_{mn,t}(q_{mn,t})=\!\!\!\!\sum\limits_{m' \in \mathcal{Q}_{m}}\!\!l_{m'n,t}(a_{m'n,t},\beta_{m'n,t}),
\vspace{-0.1cm}
\end{equation}
where $q_{mn,t}$ is a service sequence variable that satisfies $1\leqslant q_{mn,t}\leqslant\left|{\bm a_{n,t}}\right|$. $\left|{\bm a_{n,t}}\right|$~is the module of ${\bm a_{n,t}}$ and represents the number of users that are associated with HAB $n$. $\mathcal{Q}_{m}\!=\!\{m'\left|q_{m'n,t}\!<\! q_{mn,t}\right.\}$ is the set of users that are served by HAB $n$ before user $m$. Given the access delay and processing delay of each user, the total delay for user $m$ to process a computational task can be given by:
\vspace{-0.2cm}
\begin{equation}\label{eq:time}
t_{m,t}(\beta_{mn,t},a_{mn,t},q_{mn,t})\!=\!l^{\rm S}_{mn,t}(q_{mn,t})\!+\!l_{mn,t}\!\left( \beta_{mn,t},a_{mn,t}\right).
\end{equation}

\subsection{Energy Consumption Model}
In our model, the energy consumption of each user consists of three components: a) Device operation energy consumption, b) Data transmission energy consumption, and c) Data computing energy consumption. Here, the device operation energy consumption relates to the energy consumption caused by the users using their devices for any applications. The energy consumption of user $m$ can be given by \cite{MZWEC}:
\vspace{-0.2cm}
\begin{equation}\label{eq:energy}
e_{m,t}\left( \beta_{mn,t}, {a_{mn,t}}\right) = O_m\!+\!\varsigma_m \left(f^\textrm{U}_m\right)^2 \left(1\!-\! \!\beta_{mn,t}\right) z_{m,t}\!+\!P_{\rm U}l_{mn,t}^\textrm{U}\left( \beta_{mn,t},{a_{mn,t}}\right),
\vspace{-0.1cm}
\end{equation}\vspace{-0.1cm}where $O_m$ is the energy consumption of device operation and  $\varsigma_m$ is the energy consumption coefficient depending on the chip of user $m$'s device. In (\ref{eq:energy}), $\varsigma_m \left(f^\textrm{U}_m\right)^2\left(1-\beta_{mn,t}\right) z_{m,t}$ is the energy consumption of user $m$ computing the size of task $\left(1-\beta_{mn,t}\right) z_{m,t}$ at its own device and $P_{\rm U}l_{mn,t}^\textrm{U}\left( \beta_{mn,t},{a_{mn,t}}\right)$ represents the energy consumption of task transmission from user $m$ to HAB $n$.

Similarly, the energy consumption of each HAB can be given by:
\vspace{-0.2cm}
\begin{equation}\label{eq:energyHAB}
e_{n,t}\left( \beta_{mn,t}, {a_{mn,t}}\right) = O_{n}\!+\!\varsigma \left(f^\textrm{B}\right)^2  \!\beta_{mn,t} z_{m,t}\!+\!P_{\rm B}l_{mn,t}^\textrm{D}\left( \beta_{mn,t},{a_{mn,t}}\right),
\vspace{-0.1cm}
\end{equation}where $O_{n}$ is the energy consumption of hover for the HAB and $\varsigma$ is the energy consumption coefficient depending on the chip of HAB's device. In (\ref{eq:energyHAB}), $\varsigma \left(f^\textrm{B}\right)^2\beta_{mn,t} z_{m,t}$ is the energy consumption of HAB $n$ computing the data size $\beta_{mn,t} z_{m,t}$ of task that is offloaded from user $m$ and $P_{\rm B}l_{mn,t}^\textrm{D}\left( \beta_{mn,t},{a_{mn,t}}\right)$ represents the energy consumption of task transmission from HAB $n$ to user $m$.

\subsection{Problem Formulation}
We now formally pose our optimization problem whose goal is to minimize weighted sum of the energy and time consumption of each user. The minimization problem of the energy and time consumption for all users involves determining user association, service sequence, and the size of the data that must be transmitted to the HAB, as per the below formulation:
\addtocounter{equation}{0}
\begin{equation}\label{eq:max}
\begin{split}
\mathop {\min }\limits_{{\bm A}_{t},{\bm Q}_{t},{\bm \beta}_{t}}  \sum\limits_{t = 1}^T \sum\limits_{m = 1}^M \left( \gamma_{\rm E} e_{m,t}\left( \beta_{mn,t}, {a_{mn,t}}\right)\!+\!\gamma_{\rm T} t_{m,t}\!\left( \beta_{mn,t},a_{mn,t},q_{mn,t}\right)\right)
\end{split}
\end{equation}
\vspace{-0.7cm}
\begin{align}\label{c1}
\setlength{\abovedisplayskip}{-20 pt}
\setlength{\belowdisplayskip}{-20 pt}
&\;\;\;\rm{s.\;t.}\;\scalebox{1}{${a_{mn,t}} \in \left\{ {0,1} \right\},\forall {n} \in \mathcal{N},\forall {m} \in \mathcal{M},$}\tag{\theequation a}\\
&~~~~~~~\sum\limits_{n \in \mathcal{N}} a_{mn,t}\leqslant1,\forall {m} \in \mathcal{M},\tag{\theequation b}\\
&~~~~~~~1 \leqslant q_{mn,t} \leqslant\left|{\bm a_{n,t}}\right|,q_{mn,t}\in \mathbb{Z}^+,\forall m \in \mathcal{M},\forall {n} \in \mathcal{N},\tag{\theequation c}\\
&~~~~~~~q_{mn,t}\ne q_{m'n,t},\forall m\ne m', m, m' \in \mathcal{M},\forall {n} \in \mathcal{N},\tag{\theequation d}\\
&~~~~~~~0 \leqslant \beta_{mn,t} \leqslant 1,\forall m \in \mathcal{M},\forall {n} \in \mathcal{N},\tag{\theequation e}\\
&~~~~~~\sum\limits_{m = 1}^M e_{n,t}\left( \beta_{mn,t}, {a_{mn,t}}\right)  \leqslant E_t,\forall n \in \mathcal{N},\tag{\theequation f}
\end{align}
where ${\bm A}_{t}\!\!=\!\![\bm a_{1,t},\ldots,\bm a_{N,t}]$ with $\bm a_{n,t}\!\!=\!\!(a_{1n,t}, \ldots, a_{Mn,t})$,  ${\bm Q}_{t}\!\!=\!\![\bm q_{1,t},\ldots,\bm q_{N,t}]$ with $\bm q_{n,t}\!\!=\!\!(q_{1n,t}, \ldots,$\\$q_{mn,t})$, and $\bm \beta_t\!\!=\!\![\bm\beta_{1,t}, \ldots,\bm\beta_{N,t}]$ with $\bm \beta_{n,t}\!\!=\!\!(\beta_{1n,t},\ldots,\beta_{Mn,t})$. $\gamma_{\rm E}$ and $\gamma_{\rm T}$ are weighting~parameters that combine the value of energy and time consumption into an integrated utility function. (\ref{eq:max}a) and (\ref{eq:max}b) ensure that each user can connect to only one HAB for task processing. (\ref{eq:max}c) and (\ref{eq:max}d) guarantee that each HAB can only process one computational task at each time instant. (\ref{eq:max}e) indicates that the data requested by each user can be cooperatively processed by both HABs and users. (\ref{eq:max}f) is the energy constraint of HAB $n$ at time instant $t$. As the data size of the requested computational task varies, the HABs must dynamically adjust the user association, service sequence, and task allocation to minimize each user's energy and time consumption. The problem in (\ref{eq:max}) is challenging to solve by conventional optimization algorithms due to the following reasons. First, each HAB must collect the information related to the computational task requested by each user so as to minimize the energy and time consumption of ground users. However, each computational task is generated by a ground user and, hence, each HAB can only collect the information related to the computational tasks of its associated users instead of all users' computational information. When using optimization techniques, given the computational task information of only a fraction of the users, each HAB must use traditional iterative methods to find the globally optimal user association thus increasing the delay for processing computational task. Second, as the data size of each computational task varies, the HABs must re-execute the iterative methods which leads to additional delays and overhead. Thus, we need a machine learning approach that can predict the optimal user association via using the information collected by each HAB itself. Based on the predicted optimal user association, each HAB can collect the data size of the computational task from its associated users thus optimizing service sequence and task allocation for the users. User association can be considered as a multi-classification problem and SVM methods are good at solving such problems \cite{SVM}. Hence, we propose an SVM-based machine learning approach for predicting user association. In addition, exchanging the information related to historical computational task request among HABs can lead to significant energy consumption \cite{MMA}. Thus, we propose an SVM-based FL algorithm to determine the user association proactively so as to minimize the energy and time consumption. The proposed algorithm enables each HAB to use its local dataset to collaboratively train an optimal SVM model that can determine user association for all users while keeping the training data local. Based on the proactive user association, the optimization problem in (\ref{eq:max}) can be simplified and solved.



\section{Federated Learning For Proactive User Association}
Next, we introduce the training process of the SVM-based FL model for predicting user association. The proposed algorithm first enables each HAB to train an SVM model locally via using its locally collected data so as to build a relationship between each user's future association and the data size of the task that the user must process currently. Then, each HAB exchanges the trained SVM model with other HABs to integrate the trained SVM models and improve the SVM model locally so as to collaboratively perform a prediction for each user without training data exchange.

\vspace{-0.2cm}
\subsection{Components of the SVM-based FL}
An SVM-based FL algorithm consists of four components: a) agents, b) input, c) output, d) SVM model, which are defined as follows:

\begin{itemize}
\item \emph{Agents}: The agents in our system are the HABs. Since each SVM-based FL algorithm typically performs prediction for just one user, each HAB must implement $M$ SVM-based FL algorithms to determine the optimal user association for all users. Hereinafter, we introduce an SVM-based FL algorithm for the prediction of user $m$'s future association. For simplicity, an SVM-FL model of HAB $n$ is short for an SVM-FL model that HAB $n$ uses for the prediction of user $m$'s future association.

\item \emph{Input}: The input of the SVM-based FL algorithm that is implemented by HAB $n$ for predicting user $m$'s future association is defined by $\bm X_{mn}$ that includes user $m$'s user association and the data size of its requested task at historical time instants. Here, $\bm X_{mn}\!=\!\{\!(\boldsymbol{x}_{m,1},\! a_{mn,1}),\!\ldots,
    \!(\boldsymbol{x}_{m,K},\! a_{mn,K})\!\}$ where $K$ is the number of the data samples of each user $m$ collected by HAB $n$. In $\!(\boldsymbol{x}_{m,k},\! a_{mn,k}\!)$, $\boldsymbol{x}_{m,k}\!=\![x_{m,k}^{\rm X}, x_{m,k}^{\rm Y}, z_{m,k}]^{\rm T}$ with $x_{m,k}^{\rm X}$ and $x_{m,k}^{\rm Y}$ being the location of user $m$ at current time instant, $a_{mn,k}$ is the index of the user association between user $m$ and HAB $n$ at the next time instant.

\item \emph{Output}: The output of the proposed algorithm performed by HAB $n$ for predicting user $m$'s future association at time instant $t$ is $a_{mn,t+1}$ that represents the user association between HAB $n$ and user $m$ at next time instant.

\item \emph{SVM model}: For each user $m$, we define an SVM model represented by a vector $\boldsymbol{w}_{mn}$ and a matrix $\bm \Omega_m\!\in\!\mathbb{R}^{N\!\times\!N}$ where $\boldsymbol{w}_{mn}$ is used to approximate the prediction function between the input $\boldsymbol{x}_{m,t}$ and the output $a_{mn,t+1}$ thus building the relationship between the future user association and the data size of the task that user $m$ needs to process currently. $\bm \Omega_m$ is used to measure the difference between the SVM model generated by HAB $n$ and other SVM models that are generated by other HABs for determining user $m$'s future association hence improving the performance of HAB $n$'s local SVM model for prediction.
    \end{itemize}
\subsection{Training of SVM-based FL}
We must train the SVM-based FL algorithm so as accurately determine each user $m$'s association with all HABs. Training is done in a way to solve \cite{FL}:
\vspace{-0.1cm}
\begin{equation}\label{eq:12loss}
\begin{split}
&\mathop {\min }\limits_{\boldsymbol{W}_{\!m},\bm{\Omega}_{m}} \sum\limits_{n = 1}^N \sum\limits_{k = 1}^{K} \!\left\{ {l_{n}({\bm w_{mn}},(\bm x_{m,k}, a_{mn,k}))\! +\! \! \mathcal{R}(\boldsymbol{W}_{\!m},\!\bm{\Omega}_{m})} \right\},\
\end{split}
\end{equation}
\vspace{-0.7cm}
\begin{align}\label{b1}
&{\kern 1pt}{\rm s.\;t.}{\kern 5pt}\bm{\Omega}_{m} \succeq {\rm{0}}, {\kern 20pt}{\rm tr}(\bm{\Omega}_{m}) =1,\ \tag{\theequation a}
\end{align}where $l_{n} (({\boldsymbol{w}_{mn}})^{\rm T}\!\boldsymbol{x}_{m,k},\!a_{mn,k}\!)\!\!=\!\!(a_{mn,k}\!-\!(\boldsymbol{w}_{mn})^{\rm T}\boldsymbol{x}_{m,k})^2$ is a loss function that measures a squared error between the predicted user association and the target user association. $\mathcal{R}(\boldsymbol{W}_{\!m},\!\bm{\Omega}_m)\!\!=\!\!{\lambda _1}\left\| \boldsymbol{W}_m \right\|^2_F+{\lambda _2} {\rm tr}(\boldsymbol{W}_m(\bm{\Omega}_m)^{-1} ({\boldsymbol{W}_m)^{\rm T}})$ with $\lambda_1, \lambda_2>0$ is used to collaboratively build an SVM-based FL model where $\left\| \boldsymbol{W} \right\|^2_F$ is used to perform $L_2$ regularization on each local model, and ${\rm tr}(\boldsymbol{W}_m(\bm{\Omega}_m)^{-1} ({\boldsymbol{W}_m)^{\rm T}}$ captures the relationship among SVM models so as to improve the performance of SVM models that are used to determine user $m$'s association. In (\ref{eq:12loss}a), $\bm{\Omega}_m \!\succeq \!0$ implies that matrix $\bm{\Omega}_m$ is positive semidefinite.

To solve the optimization problem in (\ref{eq:12loss}), we observe the following: a) Given $\bm{\Omega}_m$, updating $\boldsymbol{W}_m$ depends on the data pair $(\boldsymbol{x}_{m,k},a_{mn,k})$ which is collected by HAB $n$ and b) Given $\boldsymbol{W}_m$, optimizing $\bm{\Omega}_m$ only depends on $\boldsymbol{W}_m$ and not on data pair $(\boldsymbol{x}_{m,k},a_{mn,k})$.
Based on these observations, it is natural to divide the training process of the proposed algorithm into two stages: a) $\bm W_m$ training stage in which HAB $n$ updates $\boldsymbol{w}_{mn}$ using its local collected data and b) $\bm \Omega_m$ training stage in which HAB $n$ first transmits $\bm w_{mn}$ to other HABs to generated $\bm W_m$ and then, calculates $\bm{\Omega}_{m}$ using its generated $\boldsymbol{W}_m$ to capture the relationship between the SVM model generated by HAB $n$ and other SVM models that are generated by other HABs for determining user $m$'s future association thus improving $\bm w_{mn}$ for each HAB $n$. Next, we introduce the two stages of the training process.


\begin{itemize}
\item \emph{$\bm W_m$ training stage}: In this stage, HAB $n$ updates $\bm w_{mn}$ based on the local dataset $\bm X_{mn}$ and $\bm{\Omega}_m$ that is calculated at last iteration. Next, we first introduce the use of quadratic approximation to divide the optimization problem in (\ref{eq:12loss}) into distributed subproblems and then, the distributed subproblems that are solved  by each HAB is presented. Given $\bm\Omega_m$, the dual problem of (\ref{eq:12loss}) can be rewritten as:
    \vspace{-0.1cm}
    \begin{equation}\label{eq:a45}
    \begin{aligned}
    \mathop {\min }\limits_{\bm \alpha_m}  \left\{ {D\left( \bm \alpha_m  \right) \!=\! \sum\limits_{n = 1}^{N}{\sum\limits_{k = 1}^{K} \!{l_{{n}}^*\! \left( { - {\alpha_{mn,k}}} \right) \!+\! {R^ * }\left({\bm X_m \bm \alpha_m \left|{\bm{\Omega}_m}\right. } \right)} } } \right\},
    \end{aligned}
    \end{equation}where $l_{n}^*(-{\alpha_{mn,k}})={\rm max}(-{\alpha_{mn,k}}\bm w_{mn}\bm x_{m,k}\!-l_{n}(\bm w_{mn}\bm x_{m,k}))$ and $R^*\left({\bm X_m \bm \alpha_m \left|{\bm{\Omega}_m}\right. } \right)={\rm max}$\\
    $(\bm X_m\bm \alpha_m \bm W_m\!-\mathcal{R}(\boldsymbol{W}_m\! \left|{\bm{\Omega}_m}\right.))$. In (\ref{eq:a45}), $\bm X_m\!\!=\!{\rm Diag}[\bm X_{m1},\!\ldots\!,\!\bm X_{mN}]$ and $\bm \alpha_m\!\!=\!\![\bm \alpha_{m1},\!\ldots\!,\!\bm\alpha_{mN}]$ where $\bm \alpha_{mn}\!\!=\!\![ \alpha_{mn,1},\!\ldots\!,\!\alpha_{mn,K}]$ with $ \alpha_{mn,k}$ being the dual variable for the data sample $({\bm x}_{m,k},a_{mn,k})$. Note that, given dual variables $\bm \alpha_{mn}$, the primal variables $\bm w_{mn}$ can be found via $\bm W_m(\bm \alpha_m)\!=\!\nabla R^*\!\left({\bm X_m \bm \alpha_m\! \left|{\bm{\Omega}_m}\right. } \right)$ where $\bm w_{mn} $ is column $n$ of $\bm W_m(\bm \alpha_m)$.

    To solve (\ref{eq:a45}) in a distributed manner, we define a local dual problem to approximate (\ref{eq:a45}). Using a quadratic approximation, this the local dual problem will be:
    \begin{equation}\label{eq:a46}
    \begin{aligned}
    &\mathop {\min }\limits_{\Delta {\bm\alpha _{mn}}}\mathcal{G}_{n}^{\sigma}\!(\!\Delta {\bm\alpha _{mn}};\!{\bm w_{mn}},\!{\bm\alpha _{mn}}\left|{\bm{\Omega}_m}\!\right.)\!\!\\
    &=\!\!\!\! \sum\limits_{k = 1}^{K}\!{{l_{n}^*}\!(\!-{\alpha_{mn,k}}\!\!-\!\! \Delta {\alpha_{mn,k}})} \!\!+\!\!\left\langle{{\bm w_{mn}}(\bm\alpha_{mn}),\!{\bm X_{mn}}\Delta {\bm\alpha_{mn}}} \right\rangle \!\!+\!\!\frac{\sigma}{2\mu_1}\!{\left\|{\bm X_{mn}}\Delta{\bm\alpha_{mn}}\right\|^2} \!\!\!+\!{R^*}\!(\bm X_n \bm\alpha_{mn}\!\left|{\bm{\Omega}_m}\!\right. ),
    \end{aligned}\vspace{-0.1cm}
    \end{equation}
    where ${\sigma}\!=\!\!\!\!\mathop {\max }\limits_{\bm\alpha_{mn}  \in {{ \mathbb{R}}^{K}}}\!\! \frac{{\left\| {{\bm X_m}\bm\alpha_{mn} } \right\|^2}}{\sum\limits_{n = 1}^N\left\| {\bm X_{mn}}\bm\alpha_{mn}  \right\|^2}\!\in\!(0,1)$ measures the correlation between each HAB's dataset that includes user $m$'s historical user association and the data size of the requested task. $\Delta\bm \alpha_{mn}\!=\![\Delta \alpha_{mn,1},\!\cdots,\!\Delta \alpha_{mn,K}]$ represents the difference between $\bm \alpha_m$ in (\ref{eq:a45}) and $\bm \alpha_{mn}$ in (\ref{eq:a46}). From (\ref{eq:a46}), we can see that, to solve the local dual problem, we only need to use the data collected by each HAB $n$. Hence, the problem in (\ref{eq:a45}) can be approximated by (\ref{eq:a46}) and solved by each HAB in a distributed manner. Note that, since a quadratic approximation is used to solve $D\left( \bm \alpha_m \right)$ in (\ref{eq:a45}), the performance loss generated by this approximation method is $D\left( \bm \alpha_m  \right)\!-\!\!\!\sum\limits_{n = 1}^N\! {\mathcal{G}_{n}^{\sigma}(\Delta {\bm\alpha_{mn}};{\bm w_{mn}},{\bm\alpha_{mn}} \left| \bm{\Omega}_m)\right.}$. In Section V, we will quantify this performance loss and show that as the number of iterations increases, the value of $D\left( \bm \alpha_m  \right)\!-\!\!\!\sum\limits_{n = 1}^N\! {\mathcal{G}_{n}^{\sigma}(\Delta {\bm\alpha_{mn}};{\bm w_{mn}},{\bm\alpha_{mn}} \left| \bm{\Omega}_m)\right.}$ decreases and thus, the solution of local dual problem in (\ref{eq:a46}) converges to the solution of global dual problem in (\ref{eq:a45}).



\item \emph{$\bm \Omega_m$ training stage}: In this stage, each HAB $n$ first transmits $\bm w_{mn}$ to other HABs and generates $\bm W_m$. Based on $\bm W_m$, each HAB $n$ calculates a structure matrix $\bm {\Omega}_m$ to measure the difference of $\bm w_{mn}$ among HABs and build an SVM model that can quantify the relationship between user association and the historical computational task information so as to predict the association result for all users. Given $\bm W_m$, (\ref{eq:12loss}) can be rewritten as:
    \vspace{-0.1cm}
    \begin{equation}\label{eq:Rloss}
    \begin{aligned}
    \begin{split}
    &\mathop {\min }\limits_{{\bm \Omega_m}}{\rm tr}(\bm W_m ({{\bm \Omega_m}})^{-1} ({\boldsymbol{W}_m)^{\rm T}}),\
    \end{split}
    \end{aligned}
    \end{equation}
    \vspace{-0.6cm}
    \begin{align}\label{b1}
    &{\kern 1pt}{\rm s.\;t.}{\kern 5pt}\bm{\Omega}_m  \succeq {\rm{0}},
    {\kern 20pt}{\rm tr}(\bm{\Omega}_m) =1.\ \tag{\theequation a}
    \end{align}
    From (\ref{eq:Rloss}), we can see that, compared to the standard FL algorithm in \cite{FLAVE} that directly averages the learning parameters $\boldsymbol{W}_m$, the proposed FL algorithm uses a matrix $\bm{\Omega}_m$ to find the relationship among all HABs' user association schemes. This approach can, in turn, improve the FL prediction performance. Given (\ref{eq:Rloss}) and (\ref{eq:Rloss}a), we have:
    \begin{equation}\label{eq:Rlosa}
    \begin{aligned}
    {\rm tr}(\boldsymbol{W}_m (\bm{\Omega})^{-1} ({\boldsymbol{W}_m)^{\rm T}})\!&=\!{\rm tr}(\boldsymbol{W}_m (\bm{\Omega}_m)^{-1}({\boldsymbol{W}_m)^{\rm T}}){\rm tr}(\bm{\Omega}_m)\\
    &\geqslant\!\!({\rm tr}(\bm{\Omega}_m)\!^{-\frac{1}{2}}\!(({\boldsymbol{W}_m)\!^{\rm T}}{\boldsymbol{W}_m})^{\frac{1}{2}}(\bm{\Omega}_m)^{\frac{1}{2}})^2\\
    &=\!\!({\rm tr}(({\boldsymbol{W}_m)^{\rm T}}{\boldsymbol{W}_m})^{\frac{1}{2}})^2,
    \end{aligned}
    \end{equation}where the inequality holds due to the Cauchy-Schwarz inequality for the Frobenius norm.
    Moreover, ${\rm tr}(\boldsymbol{W}_m\!(\bm{\Omega}_m)\!^{-1}\!({\boldsymbol{W}_m)\!^{\rm T}})$ achieves its minimum value $({\rm tr}(({\boldsymbol{W}_m)\!^{\rm T}}{\boldsymbol{W}_m})^{\frac{1}{2}})^2$ if and only if $(\bm{\Omega}_n)^{-\frac{1}{2}}(({\boldsymbol{W}_m)\!^{\rm T}}{\boldsymbol{W}_m})^{\frac{1}{2}}\!=\!a\bm{\Omega}_m$ for some constant $a$ and ${\rm tr}(\bm{\Omega}_n)\!=\!1$. Given (\ref{eq:Rlosa}), we have:
    \begin{equation}\label{eq:18loss}
    \begin{aligned}
     \bm{\Omega}_m=\frac{((\boldsymbol{W}_m)^{\rm T}\boldsymbol{W}_m)^{\frac{1}{2}}}{{\rm tr}(((\boldsymbol{W}_m)^{\rm T}\boldsymbol{W}_m)^{\frac{1}{2}})},
    \end{aligned}
    \end{equation}

   At each learning step, HAB $n$ first updates $\bm w_{mn}$ based on $\bm X_m$ and $\bm{\Omega}_m$, then broadcasts $\bm w_{mn}$ to other HABs and calculates $\bm{\Omega}_m$. Note that, the data size of $\bm w_{mn}$ can be neglected compared to the data size of each computational task and hence, the energy and time consumption for training the proposed FL is neglected. As the proposed algorithm converges, the optimal $\bm{W}_m$ and $\bm{\Omega}_m$ can be found to solve problem (\ref{eq:12loss}). The entire process of training the proposed SVM-based FL algorithm is shown in Algorithm 1.


\end{itemize}

\begin{algorithm}[t]\label{algorithm_2}
\footnotesize
\caption{Support Vector Machine Based Federated Distributed Learning Framework}
\begin{algorithmic}[1]
\STATE \textbf{Input:} Data $X_{mn}$ from $n=1,\cdots, N$ HABs, stored on one of $N$ HABs.
\STATE \textbf{Initialize:} $\boldsymbol{\Omega}_m$ is initially generated randomly via a uniform distribution. ${\bm{\alpha} ^{(0)}}: = \bm{0} \in  {\mathbb{R}^n}$.
\FOR {iterations $i=0,1,\cdots$}
\FOR {$n \in \left\{1, 2, \cdots, N \right\}$ in parallel over $N$ HABs}
\STATE For each HAB, calculating and returning $\Delta \bm{\alpha}_{mn}$ of the local subproblem in (\ref{eq:a46}).
\STATE Update local variables ${\bm{\alpha}_{mn}} \leftarrow {\bm{\alpha}_{mn}}+{\Delta \bm{\alpha}_{mn}}$.
\STATE Return updates $\bm{w}_{mn}$. 
\ENDFOR
\STATE Broadcast $\bm{w}_{mn}$ and collect trained SVM models from other HABs, save as $\bm{W}_m$.
\STATE Update $\boldsymbol{\Omega}_m$ based on $\bm{W}_m$ for latest $\bm{\alpha}_{mn}$.
\ENDFOR
\STATE \textbf{Output:} $\boldsymbol{W}_m:=\left[\bm{w}_{m1},\bm{w}_{m2},\ldots,\bm{w}_{mN}\right]$.
\end{algorithmic}
\end{algorithm}
\vspace{-0.2cm}
\section{Optimization of Service Sequence and Task Allocation}
Once the user association is determined, the HABs can optimize the service sequence and task allocation for each user so as to solve (\ref{eq:max}). Since we use directional antennas, interference among HABs is negligible. In consequence, problem (\ref{eq:max}) is independent for each HAB and can be decoupled into multiple subproblems. Given the user association, problem (\ref{eq:max}) for HAB $n$ can be rewritten as:
\vspace{-0.2cm}
\begin{equation}\label{eq:max1}
\begin{split}
&\mathop {\min }\limits_{{\bm \beta}_{n,t},{\bm s_{n,t}}} \sum\limits_{t = 1}^T \sum\limits_{m = 1}^M \left( \gamma_{\rm E} e_{m,t}\left( \beta_{mn,t}\right)\!+\!\gamma_{\rm T} t_{m,t}\!\left( \beta_{mn,t},q_{mn,t}\right)\right)
\end{split}
\end{equation}
\vspace{-0.9cm}
\begin{align}\notag
\setlength{\abovedisplayskip}{-20 pt}
\setlength{\belowdisplayskip}{-20 pt}
&\rm{s.\;t.}\;\scalebox{1}{(\ref{eq:max}c) \rule[3pt]{0.3cm}{0.05em}\;(\ref{eq:max}f)}.
\vspace{-0.2cm}
\end{align}

\vspace{-0.4cm}
Problem (\ref{eq:max1}) is a mixed integer programming problem due to the discrete variable $q_{mn,t}$ and continuous variable $\beta_{mn,t}$. To solve (\ref{eq:max1}), the following result is used so as to separate the variables $q_{mn,t}$ and $\beta_{mn,t}$ in (\ref{eq:max1}):
\vspace{-0.2cm}
\begin{lemma}
{\rm Given the data size of each computational task $z_{m,t}$, user association index $a_{mn,t}$, and service sequence variable $q_{mn,t}$, the processing delay for the users that are associated HAB $n$ will be:}
\begin{equation}\label{eq:queueproof1}
\begin{aligned}
\sum\limits_{m = 1}^{\left|{\bm a_{n}}\right|}\gamma_{\rm T} t_{mn,t}\!\left( q_{mn,t},\beta_{mn,t}\right)\!=\!\sum\limits_{m = 1}^{\left|{\bm a_{n}}\right|}\gamma_{\rm T}\left({\left|{\bm a_{n}}\right|}\!-\!q_{mn,t}\!+\!1 \right)l_{mn,t}(\beta_{mn,t}).
\end{aligned}
\end{equation}
\end{lemma}
\vspace{-0.2cm}
\begin{IEEEproof}See Appendix A.
\end{IEEEproof}
\vspace{-0.1cm}
From Lemma 1, we can see that the time needed by user $m$ and HAB $n$ to cooperatively process the computational task is determined by $\beta_{mn,t}$. We can also see that the access delay of each user $m$ is determined by $q_{mn,t}$. Next, to determine the optimal service sequence $q_{mn,t}$ of each user $m$, we state the following result:
\vspace{-0.2cm}
\begin{theorem}
{\rm {Given the data size of each computational task $z_{m,t}$ and user association index $a_{mn,t}$, the optimal service sequence of user $m$ that is associated with HAB $n$ will be $q_{mn,t}=Q_{m}$ with $Q_{m}$ being the number of users in $\mathcal{Q}_{m}\!=\!\{m'\left|l_{m'n,t}(\beta_{m'n,t})\!\leqslant\!l_{mn,t}(\beta_{mn,t})\right.\}$.
}}
\end{theorem}
\vspace{-0.3cm}
\begin{IEEEproof}See Appendix B.
\end{IEEEproof}

From Theorem 1, we can see that, the service sequence variable $\bm q_{n,t}$ can be determined according to the time consumption for processing the computational task using a sorting algorithm, such as bubble sort. Based on Theorem 1, optimization problem (\ref{eq:max1}) can be rewritten as:
\vspace{-0.2cm}
\begin{equation}\label{eq:max2}
\begin{split}
\mathop {\min }\limits_{{\bm \beta}_{n,t},{\bm q_{n,t}}}  \sum\limits_{t = 1}^T \sum\limits_{n= 1}^N \sum\limits_{m = 1}^{\left|{\bm a_{n}}\right|}\!\!\left(\gamma_{\rm E} e_{m,t}\left( \beta_{mn,t}\right)\!+\gamma_{\rm T}\left({\left|{\bm a_{n}}\right|}\!-\!q_{mn,t}\!+\!1 \right)l_{mn,t}(\beta_{mn,t})\!\right)
\end{split}
\end{equation}
\vspace{-0.7cm}
\begin{align}\label{c1}
\setlength{\abovedisplayskip}{-20 pt}
\setlength{\belowdisplayskip}{-20 pt}
&\;\;\;\rm{s.\;t.}\;\scalebox{1}{$0\leqslant \beta_{mn,t} \leqslant 1,\forall m \in \mathcal{M},\forall {n} \in \mathcal{N}$},\tag{\theequation a}\\
&~~~~~~~\sum\limits_{m = 1}^M e_{n,t}\left( \beta_{mn,t}\right)  \leqslant E_t,\forall n \in \mathcal{N}.\tag{\theequation b}
\end{align}

\vspace{-0.2cm}
Problem (\ref{eq:max2}) is a linear and convex problem since the objective functions and constraints are both convex and linear, which can be optimally solved by the well established optimization toolbox, e.g., CVX \cite{Con} optimally and efficiently.

\section{Convergence Analysis of the Proposed Algorithm}



In this section, we analyze the convergence of SVM-based FL learning algorithm and prove that the proposed FL algorithm can find the optimal user association for each HAB. To derive the convergence of SVM-based FL learning algorithm, we first show how the global dual variable $\bm \alpha_m$ in (\ref{eq:a45}) changes as the local dual variable $\bm\alpha_{mn}$ in (\ref{eq:a46}) varies, as shown in the following lemma.
\begin{lemma}
{\rm For any global dual variable} $\bm \alpha_m$, $\Delta \bm \alpha_{mn} \in \mathbb{R}^{K}$, {\rm and the learning rate} $\eta \in (0,1]$, {\rm the following relationship holds:}
\end{lemma}
\vspace{-1cm}
\begin{equation}\label{eq:a47}
\begin{aligned}
D\left( {\bm \alpha_m  \!+\! \eta {\Delta {\bm \alpha_m}} } \right) \leqslant (1 \!-\! \eta )D(\bm\alpha_m ) \!+\! \eta \!\sum\limits_{n = 1}^N {\mathcal{G}_{n}^{\sigma}(\Delta {\bm\alpha_{mn}};{\bm w_{mn}},{\bm\alpha_{mn}} \left| \bm{\Omega}_m)\right.}.
\end{aligned}\vspace{-0.1cm}
\end{equation}
\begin{IEEEproof}See Appendix C.
\end{IEEEproof}
From Lemma 2, we can see that, at each iteration, as the global dual variable $\bm\alpha_m$ is updated to $\bm\alpha_m\!+\!\eta{\Delta{\bm \alpha_m}}$, the upper bound on the value of the global dual problem $D\left( \bm\alpha_m\right)$ is formed by a sum of the values of local dual problems, $\sum\limits_{n = 1}^N\mathcal{G}_{n}^{\sigma}(\Delta {\bm\alpha _{mn}};\!{\bm w_{mn}},\!{\bm\alpha _{mn}},\!\bm{\Omega}_m)$. For each local dual problem $\mathcal{G}_{n}^{\sigma}(\Delta {\bm\alpha _{mn}};\!{\bm w_{mn}},\!{\bm\alpha _{mn}},\!\bm{\Omega}_m)$ at HAB $n$, a gradient method \cite{Convex} is used to approximate the optimal solution of the local dual problem at each iteration.
To describe the improvement of the dual objective $D\left( \bm\alpha_m\right)$ in (\ref{eq:a45}) at each iteration, we state the following proposition:




\begin{proposition}
{\rm Since $l_{n}$ that is defined in (\ref{eq:12loss}) is $(1/\mu_2)$-smooth (i.e., $l_{n}^*$ is $\mu_2$ strongly convex), there exists a constant $s \in (0,1]$, such that for any learning rate $\eta \in (0,1]$ at any iteration $h$, we have:}
\end{proposition}
\vspace{-1cm}
\begin{equation}\label{eq:a52}
\begin{aligned}
\mathbb{E}\!\left[ {D\left( \bm\alpha_m^{(h)} \right) \!-\! D\left( \bm\alpha_m^{(h + 1)} \right)} \right] \!\geqslant\! s\eta \left( {1 \!-\! \Theta } \right)\!G\left( \bm\alpha_m^{(h)} \right),
\end{aligned}
\end{equation}
where $G(\bm\alpha_m)\!=\!D(\bm\alpha_m)\!-\!\left(\!-\!\sum\limits_{n=1}^N\! {\sum\limits_{k= 1}^{K}\!\!\left\{{l_{n}((\bm w_{mn})^T{\bm x_{m,k}},{a_{mn,k}})\!+\!R(\bm W_{m},\!\bm{\Omega}_m)}\right\}\!}\right)$ is the duality gap of $\bm\alpha_m$ and $\Theta \!=\!\frac{\mathbb{E}\!\left[{\mathcal{G}_{n}^{\sigma}\!(\!\Delta \bm\alpha_{mn}^*\!;{\bm w_{mn}}\!,\bm\alpha_{mn}^{(h)}\left|{\bm{\Omega}_m}\!\right.)} -{\mathcal{G}_{n}^{\sigma}\!(\!\Delta \bm\alpha_{mn}^{(h)}\!;{\bm w_{mn}}\!,\bm\alpha_{mn}^{(h)}\left|{\bm{\Omega}_m}\!\right.)}\!\right]}{{{\mathcal{G}_{n}^{\sigma}\!(\bm0;{\bm w_{mn}}\!,\bm\alpha_{mn}^{(h)}\left|{\bm{\Omega}_m}\!\right.})}- {\mathcal{G}_{n}^{\sigma}\!(\!\Delta \bm\alpha_{mn}^*\!;{\bm w_{mn}}\!,\bm\alpha_{mn}^{(h)}\left|{\bm{\Omega}_m}\!\right.)}\!\! }$ represents the normalized distance between the local learning model obtained by the gradient method and the optimal local learning model with $\Delta\bm\alpha_{mn}^*\!=\!\! \mathop{\arg\max}\limits_{\Delta\bm\alpha_{mn}\!\in \mathbb{R}^{K}}{\mathcal{G}_{n}^{\sigma}\!(\!\Delta \bm\alpha_{mn}^{(h)};\!{\bm w_{mn}},\!\bm\alpha _{n}^{(h)},\!\bm{\Omega}_m)\!}$ being the optimal solution to the local dual problem.


\begin{IEEEproof}See Appendix D.
\end{IEEEproof}
From Proposition $1$, we can see that, the improvement of the dual objective $D(\bm \alpha_m)$ in each iteration relates to duality gap $G(\bm \alpha_m)$ and learning rate $\eta$. As learning rate $\eta$ increases, the mathematical expectation of the improvement on $D(\bm \alpha_m)$ at each iteration increases. This is because as learning rate $\eta$ increases, each local dual problem $\mathcal{G}_{n}^{\sigma}(\Delta {\bm\alpha _{mn}};\!{\bm w_{mn}},\!{\bm\alpha _{mn}},\!\bm{\Omega}_m)$ provides more information that is learned from its local dataset $\bm X_{mn}$, thus the mathematical expectation of the improvement on $D(\bm \alpha_m)$ increases. Given the relationship between the improvement of dual objective $D(\bm \alpha_m)$ and the duality gap $G(\bm \alpha_m)$ with a fixed $\eta$ at each iteration, we next estimate the change of the dual objective $D(\bm \alpha_{m})$ during $h\!+\!1$ iterations.
\begin{theorem}
{\rm Given a random initial solution $D\left(\bm\alpha_{m}^{(0)} \right)$, the gap between the optimal solution and the solution obtained after $h\!+\!1$ iterations is given by:}
\end{theorem}
\vspace{-1cm}
\begin{equation}\label{eq:a54}
\begin{aligned}
\mathbb{E}\!\left[ {\left( {D\left( {{\bm\alpha_{m}^{(h+1)}}} \right) \!-\! D\left( {{\bm\alpha_{m}^{\rm{*}}}} \right)} \right)} \right] \!\leqslant\! {\left( {1 \!-\! s\eta\! \left( {1 \!-\! \Theta } \right)} \right)^{h+1}}\!\left( {D\left( {{\bm\alpha_{m}^{(0)}}} \right) \!-\! D\left( {{\bm\alpha_{m}^{\rm{*}}}} \right)} \right).
\end{aligned}\vspace{-0.1cm}
\end{equation}
\begin{IEEEproof}See Appendix E.
\end{IEEEproof}
From Theorem 2, we observe that as the number of iteration increases, the gap between $D({\bm\alpha_{m}^{(h+1)}})$ that is obtained by the proposed algorithm and the global optimal solution $D({\bm\alpha_{m}^{\rm{*}}})$ decreases. Thus, the SVM model that is generated by all HABs will converge to a global optimal SVM model after numerous iterations. Moveover, Theorem 2 shows that, the convergence speed is affected by $\Theta$, the target accuracy of the local dual problem. As $\Theta$ decreases, at each iteration, the gap between $\mathcal{G}_{n}^{\sigma}\!(\!\Delta \bm\alpha_{mn}^*;\!{\bm w_{mn}},\!\bm\alpha_{mn}^{(h)}\left|{\bm{\Omega}_m}\!\right.)$ and $\mathcal{G}_{n}^{\sigma}\!(\!\Delta {\bm\alpha _{n}};\!{\bm w_{mn}},\!{\bm\alpha _{n}}\left|{\bm{\Omega}_m}\!\right.)$ that is obtained by gradient method decreases, and hence, the number of iterations needed for convergence decreases. Theorem 2 also shows that, as learning rate $\eta$ increases, the convergence speed increases. This is because learning rate $\eta$ affects the improvement of the dual objective $D(\bm \alpha_m)$ at each iteration, as shown in Lemma $3$. As $\eta$ increases, each local dual problem $\mathcal{G}_{n}^{\sigma}(\Delta {\bm\alpha _{mn}};\!{\bm w_{mn}},\!{\bm\alpha _{mn}},\!\bm{\Omega}_m)$ provides more information that is learned from its local dataset $\bm X_{mn}$, thus speeding up the convergence.

In terms of complexity, for training the SVM model, the major complexity in each iteration lies in finding a suboptimal solution in each HAB, which involves complexity $\mathcal{O}(N{{\rm log}_2}(1/\Theta))$ with accuracy $\Theta$ by using the gradient descent method \cite{Nuture}. Moreover, according to Theorem 1, the major complexity for optimizing the service sequence and task allocation lies in obtaining $\bm s_{n,t}$, and the complexity of calculating $\bm s_{n,t}$ depends on the sorting algorithm, such as $\mathcal{O}(M^2)$ if bubble sort is adopted.  Therefore, the proposed algorithm can run independently on each HAB due to the polynomial algorithm complexity.

\section{Simulation Results and Analysis}\label{se:system}
\begin{table}[t]
\centering
\renewcommand\arraystretch{1}
\caption{\normalsize Simulation Parameters \cite{energy}}
\small
\vspace{-0.3cm}
\setlength{\tabcolsep}{1.mm}{
\begin{tabular}{|c|c|c|c|c|c|}
\hline
\textbf{Parameter}&\textbf{Value}&\textbf{Parameter}&\textbf{Value}&\textbf{Parameter}&\textbf{Value}\\
\hline
$\emph{B}$& 10 MHz & $\varsigma$& $3.44\times10^{-23}$&$\varphi_{n,t}$&-20 dB\\
\hline
\emph{P}$_{\rm B}$& 20 W&$\omega^{\rm U}$& 1500&$\varphi_{m,t}$&-20 dB\\
\hline 
\emph{P}$_{\rm U}$& 0.5 W&$\omega^{\rm B}$& 1500&$\chi$& 1.45 dB/km\\
\hline
$\gamma_{\rm E}$& 0.5&$\emph{f}_m^{\rm U}$&0.5 GHz&$\rho$& 65\\
\hline
$\gamma_{\rm T}$& 0.5 & $\emph{f}^{\rm B}$&10 GHz&$\sigma^2$& -95 dBm\\
\hline
$\varsigma_m$& $3.44\times10^{-23}$& $f_c$&28 GHz&$H$& 17 km\\
\hline
\end{tabular}}
\vspace{-0.5cm}
\end{table}

In our simulations, an MEC-enabled HAB network area having a radius $r = 2.5$ km is considered with $M=10$ uniformly distributed users and $N=4$ uniformly distributed HABs. According to ITU guidelines \cite{ITU}, the angular variation in the location of the HABs at 17 km is less than 10$\textdegree$ for the worst-case user terminal and thus, the coverage radius of each HAB is less than 1.7 km. The values of other parameters are defined in Table I. All statistical results are averaged over 5,000 independent runs. Real data used to train the proposed algorithm is obtained from the OMNILab at Shanghai Jiao Tong University \cite{data}. We consider the data size of cellular traffic in the dataset as the data size of each user's computational task. The optimal user associations used for training the SVM model to minimize the utility function of all users are obtained by exhaustive search. In simulations, we propose two baseline algorithms named SVM-based local learning and SVM-based global learning, respectively. The SVM-based local learning enables each HAB to train its local SVM model individually while the SVM-based global learning requires each HAB to transmit its local dataset to other HABs for training purpose.
\begin{figure}[t]
\centering
\setlength{\belowcaptionskip}{-0.5cm}
\setlength{\abovecaptionskip}{-0.2cm}
\vspace{-0.2cm}
\includegraphics[width=9.3cm]{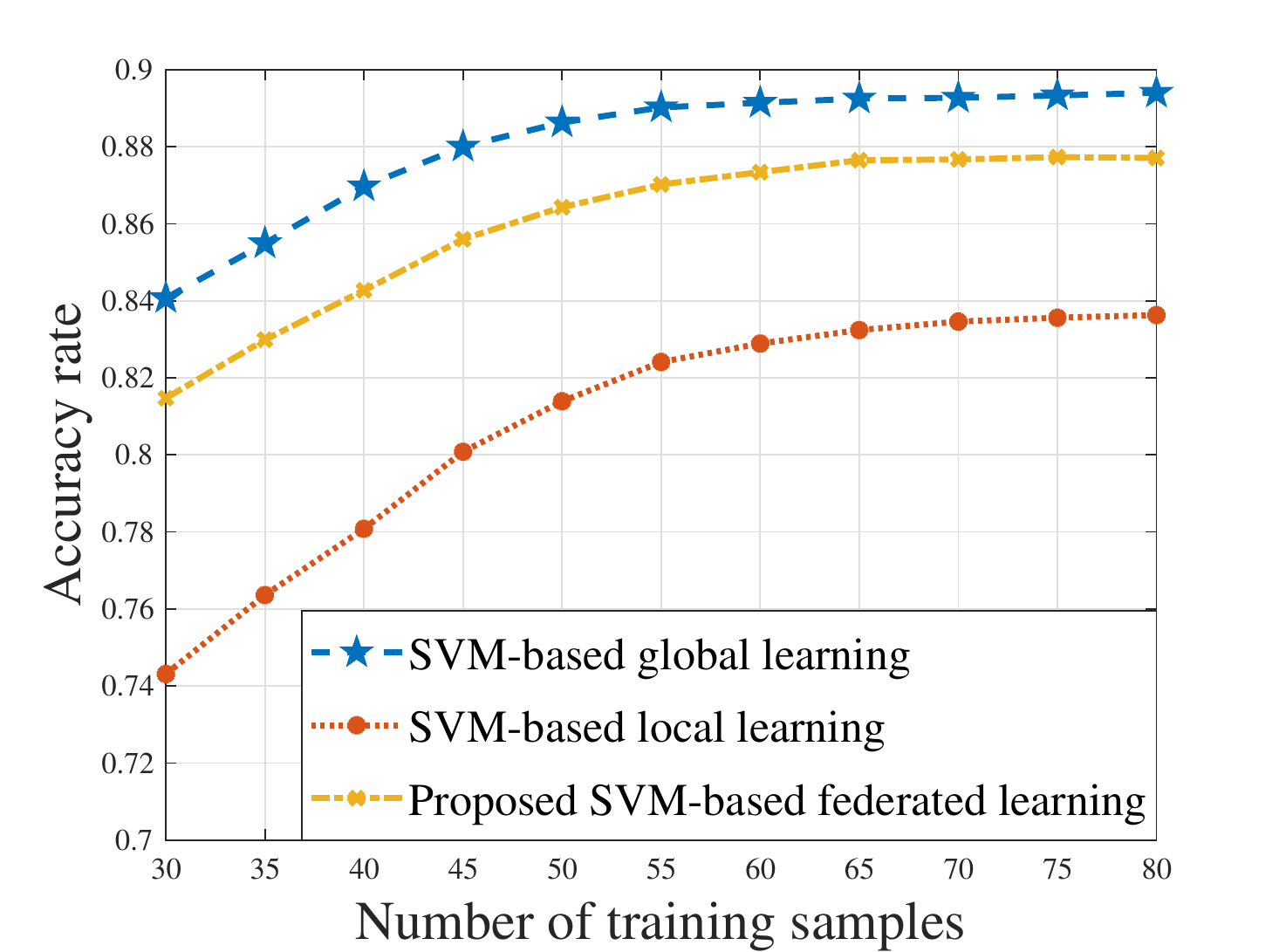}
\centering
\caption{Accuracy rate as the number of training samples varies.}
\vspace{-0.5cm}
\label{fig2}
\end{figure}

In Fig. \ref{fig2}, we show how the accuracy rate changes as the number of data samples varies. In this figure, the accuracy rate is the probability with which the considered algorithms accurately predict the optimal user association. Clearly, as the number of data samples increases, the accuracy rate of all algorithms increases. This is due to the fact that, as the number of data samples increases, the probability of underfitting decreases and hence, the accuracy rate of all considered algorithms increases. Fig. \ref{fig2} also shows that the proposed algorithm achieves only a 3\% accuracy gap compared to the SVM-based global learning. However, the SVM-based global learning algorithm requires each HAB to transmit all datasets to other HABs for training purpose, which results in high overhead as well as significant energy and time consumption for data transmission.

\begin{figure}[t]
\centering
\setlength{\belowcaptionskip}{-0.4cm}
\setlength{\abovecaptionskip}{-0.2cm}
\vspace{-0.5cm}
\includegraphics[width=9.3cm]{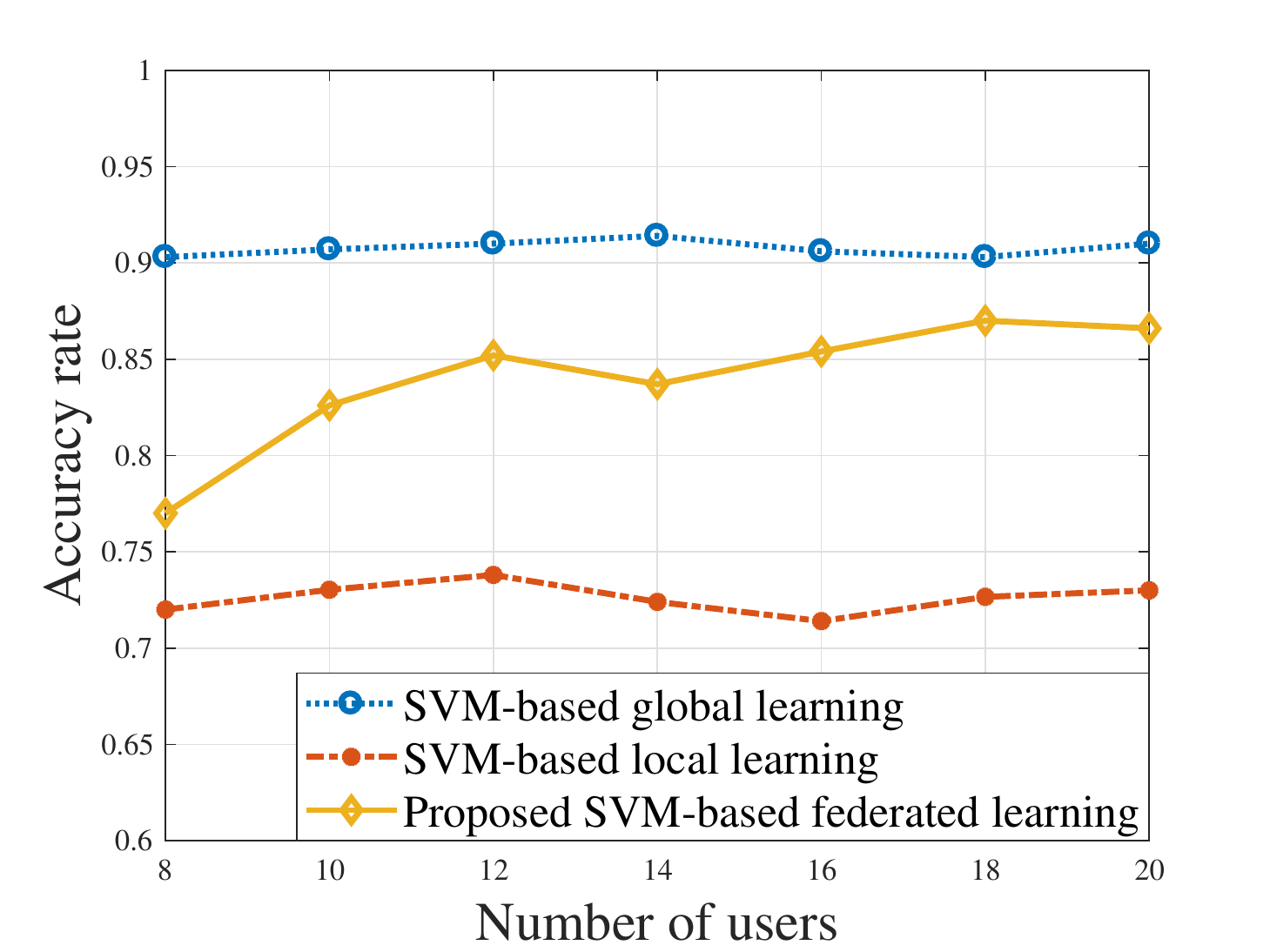}
\centering
\caption{Accuracy rate as the number of users varies.}
\vspace{-0.6cm}
\label{fig4}
\end{figure}

\begin{figure}[t]
\centering
\setlength{\belowcaptionskip}{-0.2cm}
\setlength{\abovecaptionskip}{-0.2cm}
\vspace{-0.2cm}
\includegraphics[width=9.3cm]{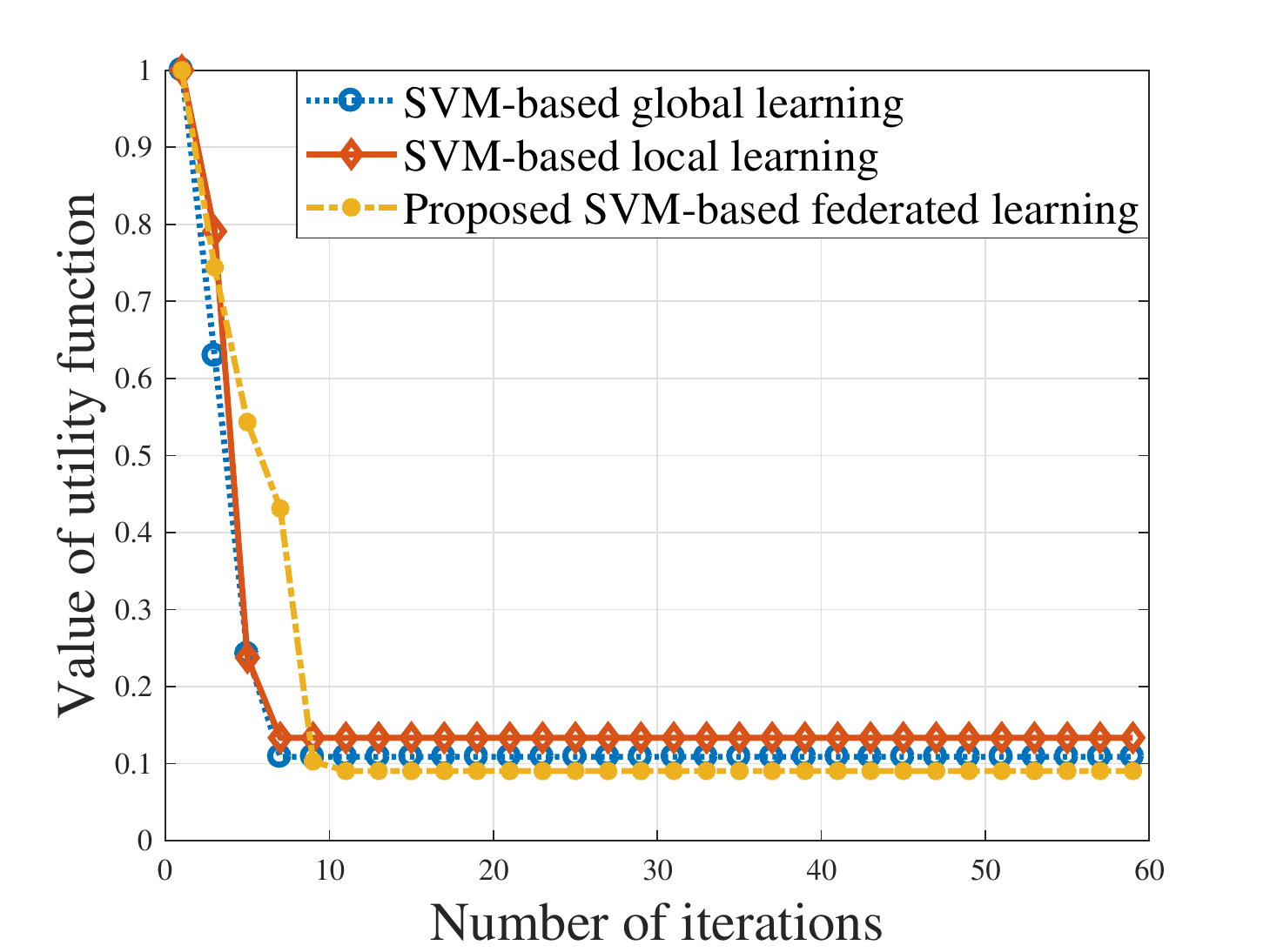}
\caption{Value of utility function as the total number of iterations varies.}
\vspace{-0.7cm}
\label{fig3}
\end{figure}

Fig. \ref{fig4} shows how the accuracy rate changes as the number of users varies. Clearly, as the number of users increases, the accuracy rate of the proposed algorithm increases. This is due to the fact that, as the number of users increases, the average energy that is used to process the computational tasks of each user decreases and hence, the probability that each user changes its association increases. In consequence, the information of computational task from each user can be collected by different HABs, thus increasing the correlation between the datasets at the HABs. Hence, the accuracy rate of the proposed algorithm increases. Fig. \ref{fig4} also shows that the proposed algorithm yields up to 19.4\% gain in terms of the accuracy rate compared to SVM-based local learning. This implies that the proposed algorithm enables each HAB to train the learning model cooperatively to build a relationship of the user association among the HABs and improve the prediction performance.





Fig. \ref{fig3} shows the number of iterations needed till convergence for all considered algorithms. From this figure, we can see that, as time elapses, the value of utility function for the considered algorithms decreases until convergence. Fig. \ref{fig3} also shows that the proposed algorithm achieves a 16.7\% loss in terms of the number of iterations needed to converge compared to the SVM-based global learning and SVM-based local learning. This is because the proposed algorithm enables each HAB to train the learning model not only based on the historical data samples, but also based on the trained parameters from other HABs, thus decreasing convergence speed. Although exchanging the trained parameters increases the number of iterations needed to converge, the proposed algorithm can achieve a performance gain of up to 19.4\% gain in terms of prediction performance compared to SVM-based local learning.


\begin{figure}[t]
\centering
\subfigure[The prediction of the user association as the data size of the computational
task varies.]{
\begin{minipage}[t]{0.5\linewidth}
\includegraphics[width=3.1in]{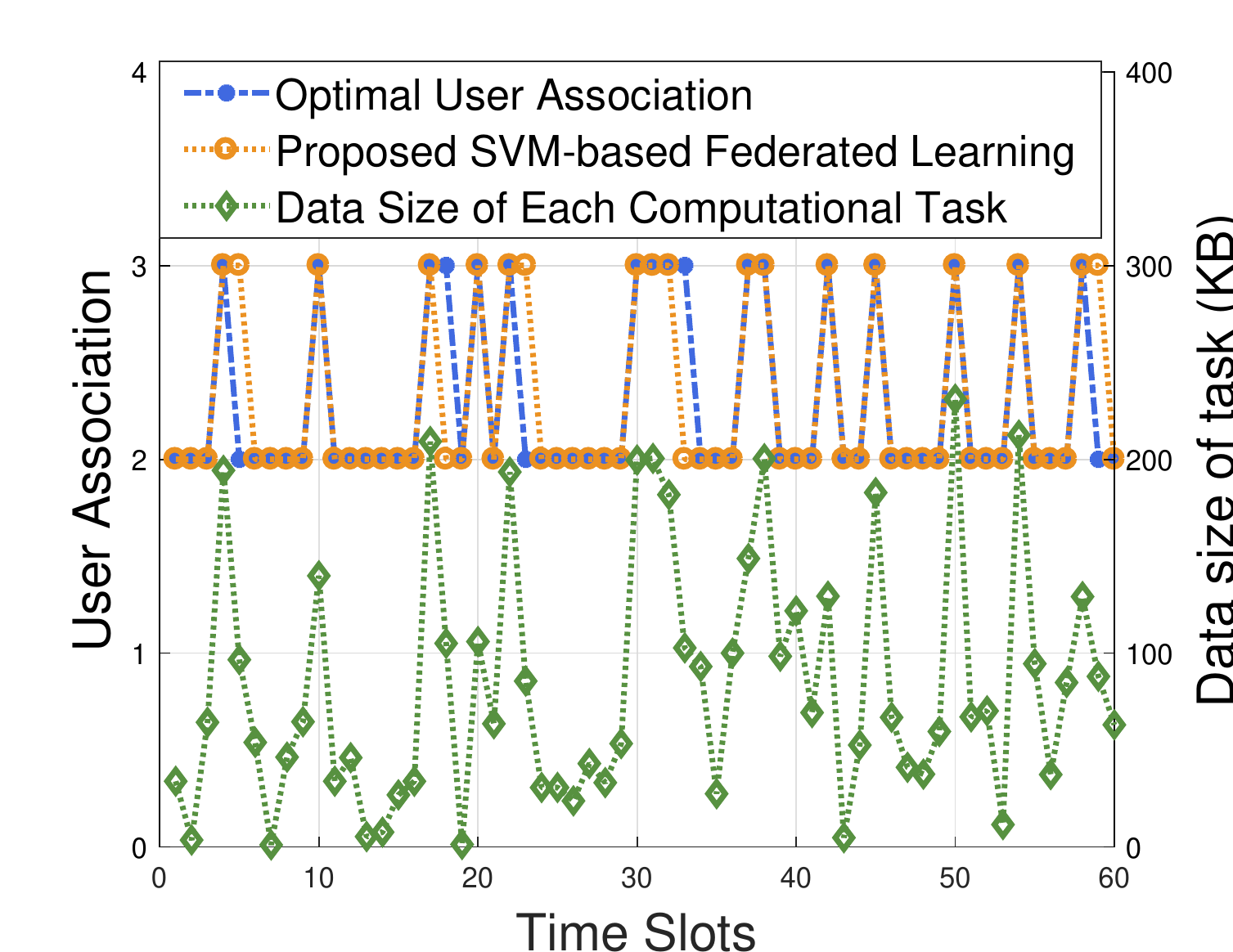}
\end{minipage}%
}%
\subfigure[The map of the distribution of the HABs and users.]{
\begin{minipage}[t]{0.54\linewidth}
\centering
\includegraphics[width=3.1in]{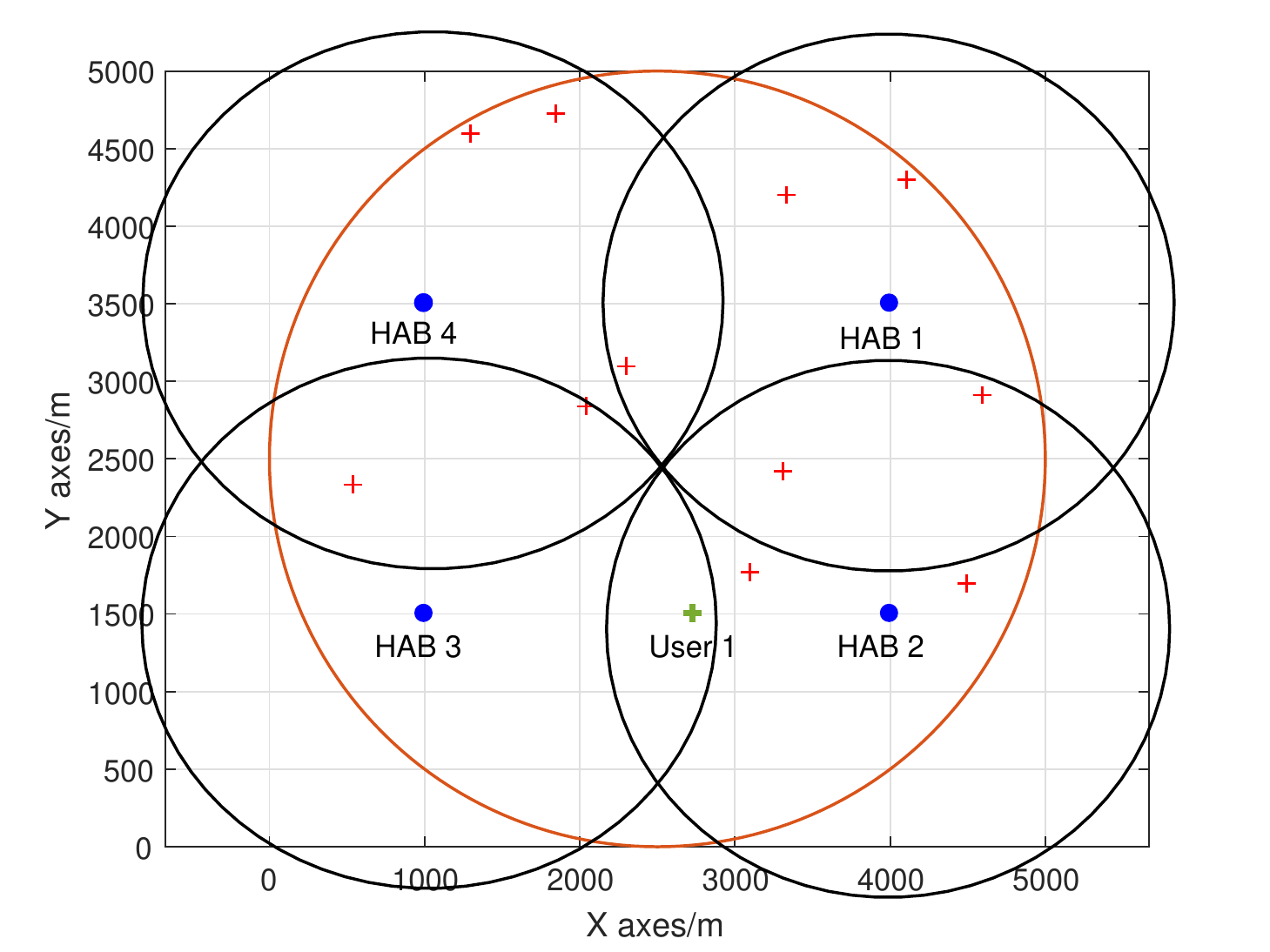}
\end{minipage}
}%
\centering
\caption{An example of the prediction of the user association performed by the proposed algorithm.}
\label{fig5}
\vspace{-0.7cm}
\end{figure}
Fig. \ref{fig5} shows an example of the prediction of the user association performed by the proposed algorithm for a network with 4 HABs and 12 users. In this figure, we can see that, as the data size of the computational task that is requested by user $1$ varies, the prediction of the user association changes, as shown in Fig. \ref{fig5}(a). This implies that the proposed algorithm enables each HAB to predict the optimal user association based on the data size of the computational task that user $1$ needs to process currently. Specifically, the proposed algorithm can achieve up to 90\% accuracy rate to predict the optimal user association. Fig. \ref{fig5}(a) also shows that user $1$ connects to HAB $3$ as long as the data size of the computational task is larger than 100 KB, and HAB $2$, otherwise. This is due to the fact that as the data size of the computational task is smaller than 100 KB, user $1$ associates with HAB $2$ for task processing since HAB $2$ is nearest to user $1$ and have enough energy to process the computational task. Moveover, as the data size of the computational task increases, each HAB needs more energy and time consumption to process the computational task that is offloaded from each user. However, from Fig. \ref{fig5}(b), we can see that, the number of users that associate with HAB $3$ is smaller than the number of users that associate with HAB $2$. Thus, as the data size of the computational task that is offloaded from user $1$ increases, the energy of HAB $2$ is insufficient to process the computational tasks from its associated users and hence, user $1$ associates HAB $3$ for task processing.

\begin{figure}[t]
\centering
\setlength{\belowcaptionskip}{-0.5cm}
\setlength{\abovecaptionskip}{-0.2cm}
\vspace{-0.1cm}
\includegraphics[width=9.3cm]{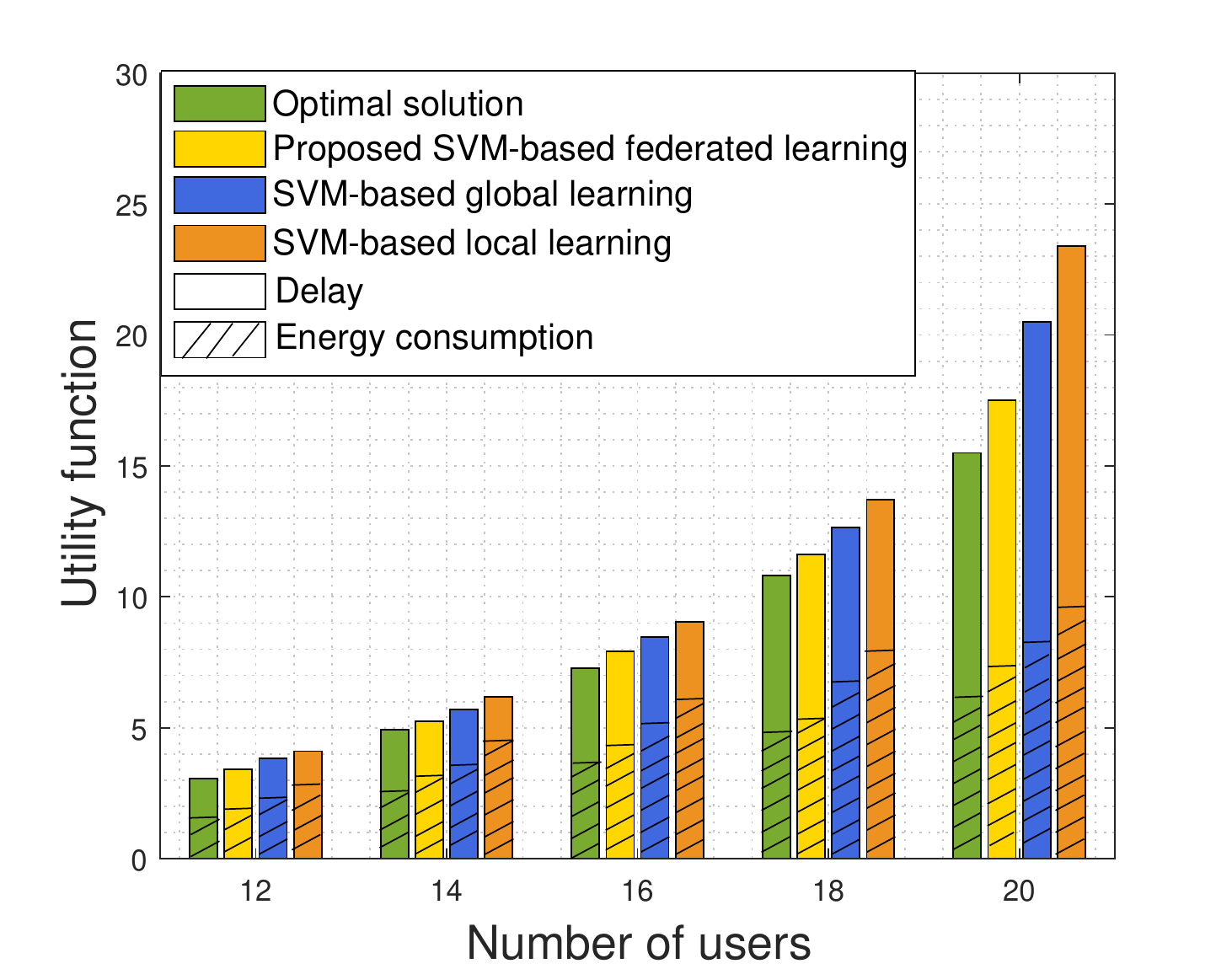}
\caption{Value of utility function as the total number of users varies.}
\vspace{-0.7cm}
\label{fig6}
\end{figure}

Fig. \ref{fig6} shows how the value of utility function changes as the number of users varies. From Fig. \ref{fig6}, we can see that the value of utility function increases as the number of users increases. This stems from the fact that, as the number of users increases, the number of tasks that users need to process increases, which increases the sum energy and time consumption for task processing. Fig. \ref{fig6} also shows that as the number of users increases, the sum energy consumption increases linearly while the sum time consumption increases exponentially. This is because that the sum energy consumption is linear related to the number of users in the considered TDMA system while the sum of the access delay is exponential related to the number of users. From Fig. \ref{fig6}, we can also see that the proposed algorithm reduces the value of utility function by up to 16.1\% and 26.7\% compared to the SVM-based global learning and SVM-based local learning. This gain stems from the fact that the proposed algorithm enables each HAB to build the SVM model cooperatively without transmitting the local training data samples to the HAB hence reducing energy consumption for local data transmission while guarantee a better performance for prediction of optimal user association.


\vspace{-0.3cm}
\section{Conclusion}
In this paper, we have studied the problem of minimizing energy and time consumption for task computation and transmission. We have formulated this problem as an optimization problem that seeks to minimize  the weighted sum of the energy and time consumption of all users. To solve this problem, we have developed an SVM-based FL algorithm which enables each HAB to cooperatively train an optimal SVM model using its own data. The SVM model can analyze the relationship between the future user association and the data size of the task that each user needs to process at current time slot so as to determine the user association proactively. Based on the optimal prediction, the optimization of service sequence and task allocation are determined so as to minimize the energy and time consumption for task computing and transmission. Simulation results have shown that the proposed approach yields significant gains in terms of sum energy and time consumption compared to conventional approaches.

\section*{Appendix}

\subsection{Proof of Lemma 1}
The enumeration method is used to prove Lemma 1.
\vspace{-0.2cm}
\begin{itemize}
  \item If the number of users associated with HAB $n$ is 1, i.e., $\left|{\bm a_{n,t}}\right|=1$, then the sum delay for processing the computational task that is requested by user $m$ is given by:
      \vspace{-0.1cm}
      \begin{equation}\notag
      \sum\limits_{m = 1}^{\left|{\bm a_{n,t}}\right|}\gamma_{\rm T} t_{mn,t}\!\left( q_{mn,t},\beta_{mn,t}\right)=\gamma_{\rm T}l^{\rm S}_{mn,t}(q_{mn,t})\!+\!\gamma_{\rm T}l_m(\beta_{mn,t})=\gamma_{\rm T}l_m(\beta_{mn,t}),
      \vspace{-0.1cm}
      \end{equation}
      where the last equality stems from the fact that the first scheduled user that is associated with HAB $n$ will finish its computational task without access delay, i.e., $l^{\rm S}_{mn,t}(q_{mn,t})\!=\!0$ with $q_{mn,t}\!=\!1$.

  \item If the number of users associated with HAB $n$ is 2, i.e., $\left|{\bm a_{n,t}}\right|\!=\!2$, then the sum delay for processing the computational tasks that are requested by the two users is given by:
      \vspace{-0.1cm}
      \begin{equation}\notag
      \begin{aligned}
      \sum\limits_{m = 1}^{\left|{\bm a_{n,t}}\right|}\!\gamma_{\rm T} t_{mn,t}\!\left( q_{mn,t},\beta_{mn,t}\right)&=\!\gamma_{\rm T}\!\left(l^{\rm S}_{m'n,t}(q_{m'n,t})\!+\!l_{m'n,t}(\beta_{m'n,t})\!+\!l^{\rm S}_{mn,t}(q_{mn,t})\!+\!l_{mn,t}(\beta_{mn,t})\right)\\ \vspace{-0.1cm}
      &=\!\gamma_{\rm T}\!\left(2l_{m'n,t}(\beta_{m'n,t})+l_{mn,t}(\beta_{mn,t})\right)
      \vspace{-0.1cm}
      \end{aligned}
      \end{equation}
      where the last equality holds since the access delay of the first scheduled user does not exist and the access delay of the second user depends on the processing delay of the first user.

  \item Using the enumeration method, if the number of users that are associated with HAB $n$ is $\left|{\bm a_{n,t}}\right|$, given the process delay for each associated users, the sum delay for processing the computational tasks that are requested by the associated users is given by:
      \vspace{-0.1cm}
      \begin{equation}\notag
      \begin{aligned}
      \sum\limits_{m = 1}^{\left|{\bm a_{n,t}}\right|}\gamma_{\rm T} t_{mn,t}\!\left( q_{mn,t},\beta_{mn,t}\right)\!=\!\sum\limits_{m = 1}^{\left|{\bm a_{n,t}}\right|}\gamma_{\rm T}\left({\left|{\bm a_{n,t}}\right|}\!-\!q_{mn,t}\!+\!1 \right)l_{mn,t}(\beta_{mn,t}).
      \end{aligned}
      \end{equation}
\end{itemize}
This completes the proof.
\vspace{-0.5cm}
\subsection{Proof of Theorem 1}
\vspace{-0.2cm}
We use contradiction method to prove Theorem 1. First, we assume that the set of users are served by HAB $n$ in an ascending order of the time consumption for processing the computational task. For a specific user $m$, the optimal service sequence is $q_{mn,t}\!=\!Q^*_{m}$ with $Q^*_{m}$ being the number of users in $\mathcal{Q}^*_{m}\!\!=\!\{m'\left|l_{m'n,t}\!\leqslant\!l_{mn,t}
\!\right.\}$. The total delay for processing can be given by:
\vspace{-0.1cm}
\begin{equation}\label{eq:queueproof1}
\begin{aligned}
\sum\limits_{m=1}^{\left|{\bm a_{n}}\right|}\!\!\gamma_{\rm T} t_{mn,t}\!=\!\gamma_{\rm T}\!\!\left(
\!\sum\limits_{m'= 1}^{Q^*_{m}\!-1}\! t_{m'n,t}\!+\!t_{mn,t}\!\left( Q^*_{m},\!\beta_{mn,t}\right)\!+\!\!\!\!\!\!\!\sum\limits_{m' = Q^*_m\!+1}^{\left|{\bm a_{n}}\right|}\!\!\!\!\!t_{m'n,t}\!\right)\!\!.
\vspace{-0.1cm}
\end{aligned}
\end{equation}
where $l^{\rm S}_{mn,t}$, $l_m$, $t_{mn,t}$ are simplified notations for $l^{\rm S}_{mn,t}(q_{mn,t})$, $l_m(\beta_{mn,t})$, and $t_{mn,t}\!\left( q_{mn,t},\beta_{mn,t}\right)$.
Then, as the optimal service sequence of the specific user $m$ is changed from $Q^*_{m}$ to $Q_{m}$, the total delay for processing can be given by:
\vspace{-0.1cm}
\begin{equation}\label{eq:queueproof2}
\begin{aligned}
&\sum\limits_{m= 1}^{\left|{\bm a_{n}}\right|}\!\!\gamma_{\rm T} t_{mn,t}
\!=\!\gamma_{\rm T}\!\!\left(\!\sum\limits_{m' = 1}^{Q_{m}\!-1}\!t_{m'n,t}\!+\!t_{mn,t}\!\left( Q_{m},\!\beta_{mn,t}\right)\!+\!\!\!\!\!\!\!\sum\limits_{m' = Q_m\!+1}^{\left|{\bm a_{n}}\right|}\!\!\!\!\!t_{m'n,t}\!\right)\!\!.
\end{aligned}
\end{equation}

\vspace{-0.1cm}
The gap between (\ref{eq:queueproof1}) and (\ref{eq:queueproof2}) is given by:
\vspace{0.1cm}
\begin{align}\notag
(\ref{eq:queueproof1})\!\!-\!\!(\ref{eq:queueproof2})\!
&=\!\gamma_{\rm T}\!\left(\sum\limits_{m' = Q_{m}}^{Q^*_m\!-1}\!\!\! t_{mn,t}\!-\!\!\!\!\!\sum\limits_{m' = Q_{m}\!+\!1}^{Q_m^*}\!\!\!\!\! t_{mn,t}\!+\!t_{mn,t}\!\left( Q^*_{m},\!\beta_{mn,t}\right)\!-\!t_{mn,t}\!\left( Q_{m},\!\beta_{mn,t}\right)\!\!\right)\\ \notag
&=\!\gamma_{\rm T}\!\left(\sum\limits_{m' = Q_{m}}^{Q^*_m\!-1}\!\!\!\!\! t_{m'n,t}\!-\!\!\!\!\!\sum\limits_{m' = Q_{m}\!}^{Q_m^*\!-1}\!\!\!\! t_{m'n,t}\!\left( q_{m'n,t}\!-\!1\right)\!+\!t_{mn,t}\!\left( Q^*_{m},\!\beta_{mn,t}\right)\!-\!t_{mn,t}\!\left( Q_{m},\!\beta_{mn,t}\right)\!\!\right)\\ \notag
&=\!\gamma_{\rm T}\!\left(\sum\limits_{m' = Q_{m}}^{Q^*_m\!-1}\!\!\!\!\left({\left|{\bm a_{n}}\right|}\!-\!q_{m'n,t}\!+\!1 \right)l_{m'n,t}\!-\!\!\!\!\!\sum\limits_{m' = Q_{m}}^{Q^*_m\!-1}\!\!\!\left({\left|{\bm a_{n}}\right|}\!-\!q_{m'n,t}\!-\!1\!+\!1 \right)l_{m'n,t}\!+\! (Q^*_m\!-\!Q_{m})
l_{mn,t}\!\right)\\ \notag
&=\!\gamma_{\rm T}\!\left(\sum\limits_{m'=Q_m}^{Q^*_m\!-1}\!\!\!l_{m'n,t}\!-\!(Q^*_m\!-\!Q_{m})
l_{mn,t}\!\right).
\end{align}

\begin{itemize}
  \item If $Q_m^*\!>\!Q_m$, user $m$ is served before the users whose required service time is less than user $m$, i.e., $l_{m'n,t}\!<l_{mn,t}$, we have: $\!\sum\limits_{m'=Q_m}^{Q_m^*\!-\!1}\!\!\!l_{m'n,t}\!-\!(Q_m^*\!-\!Q_{m})l_{mn,t}<0$.

  \item If $Q^*_m\!<\!Q_m$, user $m$ is served after the users whose required service time is larger than user $m$, i.e., $l_{m'n,t}\!>l_{mn,t}$, we have $\sum\limits_{m'=Q_m}^{Q^*_m\!-\!1}\!\!\!\!\!l_{m'n,t}\!-\!(Q^*_m\!-\!Q_{m})l_{mn,t}\!=\!
      -\!\!\!\!\!\!\!\sum\limits_{m'=Q_m^*\!-\!1}^{Q_m}\!\!\!\!\!\!\!l_{m'n,t}\!+\!(Q_m\!-\!Q^*_{m})l_{mn,t}\!<\!0$\! .
\end{itemize}

\vspace{0.1cm}
From the above analysis, we can see that, as the service sequence $Q^*_{m}$ for user $m$ changes, the time needed by all associated users for processing the computational task $\sum\limits_{m= 1}^{\left|{\bm a_{n}}\right|}\!\!\gamma_{\rm T} t_{mn,t}\!$ increases. Thus, the sum delay is minimized as the associated users are served in ascending order of the time consumption for processing the computational task, which can be expressed as the service sequence $q_{mn,t}=Q_{m}$ with $Q_{m}$ being the number of elements in $\mathcal{Q}_{m}\!=\!\{m'\left|l_{m'n,t}\!\leqslant\!l_{mn,t}\right.\}$.

\noindent This completes the proof.

\vspace{-0.4cm}
\subsection{Proof of Lemma 2}
\vspace{-0.1cm}
To prove Lemma $2$, we need to define the following functions:
\vspace{-0.3cm}
\begin{definition}
\rm{\textbf{\emph{L}-Lipschitz continuous function.}} A function $f$: $\mathbb{R}^m\rightarrow\mathbb{R}$ is \emph{L}-Lipschitz continuous if $\forall \bm a,\bm b \in \mathbb{R}^m$, we have $\left|{f(\bm a) \!-\! f(\bm b)}\right| \leqslant L\left\| {\bm a \!-\! \bm b} \right\|.$
\end{definition}
\vspace{-0.6cm}
\begin{definition}
\rm{\textbf{($\bm 1 \bm/ \bm\mu$)-smooth function.}} A function $f$: $\mathbb{R}^m\rightarrow\mathbb{R}$ is $(1/\mu)$-smooth if it is differentiable and its derivative is $(1/\mu)$-Lipschitz continuous or equivalently, i.e., $\forall \bm a,\bm b \in \mathbb{R}^m$, we have $f\left(\bm a \right) \leqslant f\left(\bm b\right) \!+\! \left\langle {\nabla f\left(\bm b\right),\bm a \!-\! \bm b} \right\rangle  \!+\! \frac{1}{{2\mu }}{\left\| {\bm a \!-\! \bm b} \right\|^2}.$
\end{definition}
\vspace{-0.3cm}

According to definition of $D(\bm \alpha)$ in (\ref{eq:a45}), we have:
\begin{align}\label{eq:app1}
D\left(\bm\alpha_m\!+\! \eta \Delta{\bm\alpha_m} \right)\! =\! \underbrace {\sum\limits_{n = 1}^N \!\left( {\sum\limits_{k = 1}^{{K}}{l_{n}^*\left( { - {\alpha_{mn,k}}\!\!-\!\eta \Delta {\alpha_{mn,k}}} \right)} \!} \right)}_{\emph{\rm the updates in } l_{n}^*} \!+\! \underbrace {R^*\!\left( {\bm X_m\!\left({\bm\alpha_m \!+\!\eta{\Delta {\bm\alpha_m}} } \right)}\!\left|{\bm{\Omega}_m}\!\right. \right)}_{\emph{\rm the updates in } R^*}\!.
\end{align}\vspace{-0.5cm}

Now, we separate the variables $\alpha_{mn,k}$ with $\Delta {\alpha_{mn,k}}$ in $l_n^*$ and $\bm \alpha_m$ with $\Delta \bm{\alpha}_m$ in $R^*$, respectively. Rewrite the updates in $l_{n}^*$ as:
\vspace{0.2cm}
\begin{align}\label{eq:app2}
&{\sum\limits_{n = 1}^N \!\left( {\sum\limits_{k = 1}^{{K}}\!{l_{n}^*\left( { - {\alpha_{mn,k}}\!\!-\!\eta \Delta {\alpha_{mn,k}}} \right)} \!} \right)}\!\!
\leqslant \!\!\sum\limits_{n=1}^N \!\!{\left( {\sum\limits_{k \!=\! 1}^{{K}}\!(1 \!-\! \eta)l_{n}^*\!\left(\!-\alpha_{mn,k} \right)\!+\!\!\!\sum\limits_{k \!=\! 1}^{{K}}\!\eta l_{n}^*\!\!\left(\!-\alpha_{mn,k}\!\!-\!\!\Delta {\alpha_{mn,k}} \right)}\!\! \right)}\!,
\end{align}where the inequality stems from the Jensen's inequality. The updates in $R^*$ can be rewritten as:
\begin{align}\label{eq:app3}
&{R^*}\!\left( (\bm X_m\bm\alpha_m\!+\! \eta{\bm X_m\Delta \bm\alpha_m}) \left|{\bm{\Omega}_m}\!\right. \right) \notag \\
&\leqslant\! {R^*}\!\left( {\bm X_m\bm \alpha_m}\left|{\bm{\Omega}_m}\!\right. \right)\!\!+\!\!\! \sum\limits_{n = 1}^N \!{\eta \nabla {R^*}\!\left( {\bm{X}_{mn}\Delta {\bm\alpha _{mn}}\left|{\bm{\Omega}_m}\!\right.} \right) \cdot } \bm{X}_{mn}\Delta {\bm\alpha_{mn}}\!\!+\!\! \frac{\eta}{2\mu_1}{\sum\limits_{n= 1}^N \!{\left\| {\bm{X}_{mn}\Delta {\bm\alpha_{mn}}} \right\|} ^2}\notag \\
& =\! {R^*}\!\left( {\bm X_m\bm\alpha_m \left|{\bm{\Omega}_m}\!\right.} \right) \!+\!\! \sum\limits_{n = 1}^N \!{\eta \left\langle {{\bm w _{mn}}(\bm\alpha_{mn} ),{\bm X}_{mn}\Delta {\bm\alpha _{mn}}} \right\rangle }  \!+\! \frac{\eta}{2\mu_1}{\sum\limits_{n = 1}^N \!{\left\| {\bm{X}_{mn}\Delta {\bm\alpha_{mn}}} \right\|} ^2} \notag\\
& \leqslant\! {R^*}\!\left( {\bm X_m\bm\alpha_m }\left|{\bm{\Omega}_m}\!\right. \right) \!+\!\! \sum\limits_{n = 1}^N {\eta \left\langle {{\bm w_{mn}}(\bm\alpha_{mn}),\bm{X}_{n}\Delta {\bm\alpha_{mn}}} \right\rangle } \!+\! \frac{{{\eta}\sigma}}{2\mu_1}{\sum\limits_{n = 1}^N\! {\left\| {\bm{X}_{mn}\Delta {\bm\alpha_{mn}}} \right\|} ^2},
\end{align}
where the first inequality stems from Definition $2$ and the second inequality stems from the definition of $\sigma$ in (\ref{eq:a46}). Substituting (\ref{eq:app2}) and (\ref{eq:app3}) into (\ref{eq:app1}), we have:
\begin{align}\label{eq:app4}
D\!\left(\bm\alpha_m\!\!+\!\eta\Delta \bm\alpha_m  \right)\!
&\leqslant\!\sum\limits_{n = 1}^N \!\!{\left(\sum\limits_{k = 1}^{K} \!\!{(\!1 \!-\! \eta )l_{n}^*\!\!\left(\!-\alpha_{{mn},k}\right)\!\!+\! \eta l_{n}^*\!\!\left(\!-\alpha_{mn,k}\!\!-\!\! \Delta {\alpha_{mn,k}} \right)}\!\! \right)}\!+\!{R^*}\!\!\left( {\bm X_m\bm\alpha_m }\!\left|{\bm{\Omega}_m}\!\right. \right)\! \notag \\
&{\kern 12pt}+\!\!\!\sum\limits_{n = 1}^N \!{\eta\! \left\langle {{\bm w_{mn}}(\bm\alpha_{mn}),\!{\bm X_{mn}}\Delta {\bm\alpha _{mn}}}\right\rangle} \!+\! \frac{\eta\sigma}{2\mu_1}\!{\sum\limits_{n = 1}^N\!{\left\| {{\bm X_{mn}}\Delta {\bm\alpha_{mn}}}\right\|}^2} \notag \\
&=\!\eta\!\!\sum\limits_{n = 1}^N\!\!\left(\! {\sum\limits_{k = 1}^{K}\!l^*\!(\!-{\alpha_{mn,k}}\!\!-\!\!\Delta {\alpha_{mn,k}}\!)\!\!+\!\! \left\langle \bm w_{mn}(\bm\alpha_{mn}),\!{\bm X_{mn}}\Delta {\bm\alpha _{mn}} \right\rangle \!\!+\! \frac{{\sigma}}{2\mu_1}\!{{\left\| {{\bm X}_{mn}\Delta {\bm\alpha _{mn}}} \right\|}^{2}}\!\!}\right) \notag \\
&{\kern 12pt}+\!\eta{R^*}\!(\bm X_{m}\bm\alpha_{m}\!\left|{\bm{\Omega}_m}\!\right.)\!\!+\!\!\left(\!1\!\!-\!\!\eta \right)\!\!\left(\!\! {R^*\!\left(\bm X_m\bm\alpha_m\! \left|{\bm{\Omega}_m}\!\right. \right)\!\!+\!\!\!\sum\limits_{n = 1}^N \!{\sum\limits_{k = 1}^{{K}}\! {l_{n}^*\!\left( \!-{\alpha_{mn,k}}\! \right)} } } \!\!\right) \notag \\
& = \!(1 \!-\! \eta )D(\bm\alpha_m ) \!+\! \eta\! \sum\limits_{n = 1}^N \!{\mathcal{G}_{n}^{\sigma}\!(\!\Delta {\bm\alpha_{mn}};\!{\bm w_{mn}},\!{\bm\alpha _{mn}}\!\left|{\bm{\Omega}_m}\!\right.)}.
\end{align}
This completes the proof.
\vspace{-0.3cm}
\subsection{Proof of Proposition 1}
\vspace{-0.2cm}
To prove Proposition $1$, we need to define the following function:
\vspace{-0.2cm}
\begin{definition}
\rm{\textbf{$\bm\mu$-strong convex function.}} A function $f$: $\mathbb{R}^m\rightarrow\mathbb{R}$ is $(\mu)$-strong convex if $\forall \bm a,\bm b \in \mathbb{R}^m$ and $\forall s \in \partial f(\bm b)$, we have $
f\left(\bm a \right) \geqslant f\left(\bm b\right) \!+\! \left\langle {s,\bm a\!-\! \bm b} \right\rangle  \!+\! \frac{\mu}{{2}}{\left\| {\bm a \!-\! \bm b} \right\|^2}$ where $\partial f(\bm b)$ represents the subdifferential of $f$ at $\bm b$.
\end{definition}
\vspace{-0.2cm}
Using the definition of the dual update $\bm \alpha_m^{(h+1)}\!=\!\bm \alpha_m^{(h)}\!+\!\eta\Delta \bm\alpha_m^{(h)}$ from Algorithm 1, we have:
\begin{align}\label{eq:app5}
&\mathbb{E}\left[{D\!\left( \bm\alpha_m^{\left(h\right)} \right) \!-\! D\!\left( \bm\alpha_m^{\left(h+1\right)}\right)}\right] \\ \notag
=& \mathbb{E}\!\left[ {D\!\left( \bm\alpha_m^{\left( h \right)} \right) \!-\! D\!\left( \!\bm\alpha_m^{\left( h \right)} \!+\! \eta {\Delta \bm\alpha^{(h)}}\right)} \right] \\ \notag
\geqslant& \mathbb{E}\!\left[ {D\!\left(\bm\alpha_m^{\left( h \right)} \right) \!-\! (1 \!-\! \eta )D({\bm\alpha_m^{\left( h \right)}}) \!-\! \eta\! \sum\limits_{n = 1}^N{\mathcal{G}_{n}^{\sigma}(\Delta\bm\alpha _{mn}^{(h)};{\bm w_{mn}},\bm\alpha_{mn}^{(h)}\left|{\bm{\Omega}_m}\!\right.)} } \right]\\ \notag
=&\eta \mathbb{E}\!\left[ {D\!\left( \bm\alpha_m^{\left( h \right)} \right) \!-\!\!\! \sum\limits_{n = 1}^N {\mathcal{G}_{n}^{\sigma}(\Delta \bm\alpha _{mn}^{(h)};{\bm w_{mn}},\bm\alpha_{mn}^{(h)}\left|{\bm{\Omega}_m}\!\right.)} } \right] \\ \notag
=&\eta \mathbb{E}\!\!\left[\!{D\!\left(\bm\alpha_m^{\left( h \right)} \right) \!\!-\!\!\!\sum\limits_{n = 1}^N\!\!\Bigg(\!\!{\mathcal{G}_{n}^{\sigma}(\Delta \bm\alpha_{mn}^*)}\!\!+\! {\mathcal{G}_{n}^{\sigma}(\Delta \bm\alpha_{mn}^*)}\!\!-\!{\mathcal{G}_{n}^{\sigma}(\Delta \bm\alpha _{mn}^{(h)};\!{\bm w_{mn}},\!\bm\alpha_{mn}^{(h)}\!\left|{\bm{\Omega}_m}\!\right.)}}\!\!\Bigg)\!\right] \\ \notag
\geqslant&\eta \mathbb{E}\!\Bigg[\!D\!\left(\bm\alpha_m^{\left( h \right)} \right) \!\!-\!\!\!\sum\limits_{n = 1}^N\!{\mathcal{G}_{n}^{\sigma}(\Delta \bm\alpha_{mn}^*)} \!\!+\!\Theta\!\Bigg(\!\sum\limits_{n = 1}^N\!{\mathcal{G}_{n}^{\sigma}(\Delta \bm\alpha_{mn}^*)}\!\!-\!\!\sum\limits_{n = 1}^N{\!{\mathcal{G}_{n}^{\sigma}(\bm0;\!{\bm w_{mn}},\!\bm\alpha_{mn}^{(h)}\!\left|{\bm{\Omega}_m}\!\right.)}}\!\!\Bigg)\! \Bigg]\\ \notag
=&\eta (1 \!-\! \Theta )\!{\Bigg(\!\!{D\!\left( {{\bm \alpha_m ^{\left( h \right)}}} \right)\!\!-\!\sum\limits_{n = 1}^N \!{\mathcal{G}_{n}^{\sigma}(\Delta \bm \alpha_{mn}^*)}\!} \Bigg)}.
\end{align}
where ${\mathcal{G}_{n}^{\sigma}(\Delta\bm\alpha _{mn}^{*})}$ is simplified notations for ${\mathcal{G}_{n}^{\sigma}(\Delta\bm\alpha _{mn}^{*};{\bm w_{mn}},\bm\alpha_{mn}^{(h)}\left|{\bm{\Omega}_m}\!\right.)}$.

Now, we derive a lower bound for the gap between the solution of global dual problem and the sum of the solutions of local dual problems. Substituting $s(\pi_k\!-\!\alpha_{mn,k})=\Delta\alpha_{mn,k}^{*}$ into (\ref{eq:app5}) yields:
\begin{align}\label{eq:app6}
&{D\!\left( {{\bm \alpha_m ^{\left( h \right)}}} \right)\!\!-\!\!\!\sum\limits_{n = 1}^N \!{\mathcal{G}_{n}^{\sigma}(\Delta \bm \alpha_{mn}^*;\!{\bm w_{mn}},\!\bm \alpha_{mn}^{(h)}\!\left|{\bm{\Omega}_m}\!\right.)}\!} \notag \\
&\!\!=\!\!\sum\limits_{n = 1}^N\!\!\left(\!{\sum\limits_{k\!=\! 1}^{{K}}\!\!{\left({l_{n}^*\!\left(\!{{-\alpha_{mn,k}}}\! \right)\!-\! l_{n}^*\!(\!-{\alpha_{mn,k}}\!\!-\!\!\Delta {\alpha_{mn,k}^{*}})} \right)\!-\! \left\langle {{\bm w_{mn}}(\bm\alpha_{mn}),\!{\bm X_{mn}}\Delta \bm\alpha_{mn}^*} \right\rangle \!\!-\!\! \frac{{\sigma}}{2\mu_1}{{\left\|{\bm{X}_{mn}\Delta \bm\alpha_{mn}^*} \right\|}^2}}}\!\!\right) \notag \\
&\!\!=\!\!\sum\limits_{n = 1}^N\!\!\left(\!\sum\limits_{k\!=\! 1}^{K} \!\!\left({l_{n}^*\!\left(\!-\alpha_{mn,k}\!\right)\!-\!\! l_{n}^*\!(\!- s{\pi_k}\!\!-\!\!(1\!\!-\!\!s){\alpha_{mn,k}})} \right)\!\!-\!\!\left\langle {{\bm w_{mn}}\!(\!\bm\alpha_{mn}\!)\!,\!s{\bm X}_{mn}\!\left({\!\bm \pi\!\!-\!\bm\alpha_{mn}}\right)} \right\rangle   \!\!-\!\frac{{\sigma}}{2\mu_1}\!{{\left\| s{\bm X_{mn}\!\left(\!{\bm \pi\!\!-\!\!{\bm\alpha _{mn}}\! }\right)} \right\|}^2}\!\!\right)\notag \\
&\!\!\geqslant\!\!\!\sum\limits_{n = 1}^N\!\!\left(\!\sum\limits_{k\!=\!1}^{{K}} \!\!\!s\!\left({l_{n}^*\!\!\left(\!-\alpha_{mn,k}\!\right)\!-\! \!l_{n}^*\!(\!{-\pi_k}\!)}\!\right)\!\!+\!\!\frac{{\mu_2\!(\!1\!\!-\!\! s\!)}}{2}\!s{{\left(\!{\bm \pi\!\!-\!\!{\bm\alpha_{mn}}}\!\right)}^2}\!\!\!-\!\! \left\langle{\bm{w}_{mn}\!(\!\bm\alpha_{mn}\!)\!,\!s{\bm X_{mn}}\!\left(\!{\bm{\pi}\!\!-\!{\bm\alpha_{mn}}} \!\right)} \right\rangle
\!\!-\! \frac{{\sigma}}{2\mu_1}\!{{\left\|s{\bm X_{mn}}\!\!\left(\!\bm \pi\!\!-\!{\bm\alpha_{mn}}\! \right) \right\|}\!^2} \!\!\right)\!\!,
\end{align}
where the inequality stems from the $\mu_2$-strong convexity of $l^*$. According to the definition of the function in (\ref{eq:a45}), we have:
\vspace{-0.3cm}
\begin{equation}\label{eq:app7}
\begin{aligned}
l_{n}^*( - {\pi}_k)\!=\!- {\pi_k}\bm w_{mn}{\left( \bm\alpha_{mn}\right)\!^{\rm T}}{\bm x_{m,k}}\!-\!l_{n}\!\left( { \bm w_{mn}{{\left(\!\bm\alpha_{mn}\right)}\!^{\rm T}}{\bm x_{m,k}}}\!\right)\!.
\end{aligned}\vspace{-0.1cm}
\end{equation}
Moreover, based on the definition of the primal and dual optimization problems in (\ref{eq:12loss}) and (\ref{eq:a45}), the duality gap $G\!\left(\bm\alpha_m\right)$ can be given by:
\begin{equation}\label{eq:app10}
\begin{aligned}
G\!\left(\bm\alpha_m\right) \!&= D(\bm\alpha_m )\!-\!\left(\!-\!\sum\limits_{n=1}^N\! {\sum\limits_{k= 1}^{K}\!\!\left\{{l_{n}((\bm w_{mn})^T{\bm x_{m,k}},{a_{mn,k}})\!+\!R(\bm W_m,\!\bm{\Omega}_m)}\right\}\!}\right)\\
&\!= \!\!\sum\limits_{n = 1}^N\!\!{\left( {\sum\limits_{k = 1}^{{K}}\!\!{\left({l_{n}^*\!\!\left(\!- {\alpha_{mn,k}} \right)\!\!+\!{l_{n}}\!(\!\bm w_{mn}{{\left(\bm\alpha_{mn}\right)}\!^{\rm T}}\!{\bm x_{m,k}} )} \right)} }\!\!\right)} \!\!\!+\! {\left( {\bm X_m\bm \alpha_m} \right)\!^{\rm T}}\bm W_m\!\left(\bm\alpha_m\right) \\
&=\!\!  \sum\limits_{n = 1}^N \!{\left({\sum\limits_{k = 1}^{{K}} \!\!{\left({l_{n}^*\!\left( {\!- {\alpha_{mn,k}}} \right)\!\!+\! {l_{n}}(\bm w_{mn}{{\left(\bm\alpha_{mn}\right)}\!^{\rm T}}{\bm x_{m,k}} )} \right)\!\!+\! {\bm\alpha_{mn}}{\bm X_{mn}}\bm w_{mn}\!\left( \bm\alpha_{mn}\right)}\! } \right)}.
\end{aligned}
\end{equation}
Substituting (\ref{eq:app10}) and (\ref{eq:app7}) into (\ref{eq:app6}), we have:
\begin{align}\label{eq:app9}
&{D\!\left( {{\bm \alpha_m^{\left( h \right)}}} \right)\!\!-\!\!\sum\limits_{n = 1}^N \!{\mathcal{G}_{n}^{\sigma}(\Delta \bm \alpha_{mn}^*;\!{\bm w_{mn}},\!\bm \alpha_{mn}^{(h)}\!\left|{\bm{\Omega}_m}\!\right.)}\!}  \notag \\
=&\! \!\sum\limits_{n = 1}^N\!\!\left(\! {\sum\limits_{k = 1}^{K}\! \!\left(\!{sl_{n}^*\!\left( {\!- {\alpha_{mn,k}}} \right) \!+\! s{\bm\alpha _{mn}} {{\bm X_{mn}}\bm w_{mn}\!\left(\bm\alpha_{mn}\right)} \!+\! s{l_{n}} (\bm w_{mn}{{\left( \bm\alpha_{mn}\right)}\!^{\rm T}}{\bm x_{m,k}})} \right) \!\!+\! \frac{{\mu_2 (1\! -\! s)}}{2}s{{\left( {\bm \pi\! \!-\! \bm\alpha_{mn}} \right)}^2}\!}\right. \notag \\
&{\kern -12pt}\left.{\!-\!\left\langle {\bm w_{mn}}\!(\bm\alpha_{mn})\!,\!s{\bm X_{mn}}\!\!\left(\!{{\bm \pi}\!\!-\!\bm\alpha_{mn}}\right) \right\rangle\!\!-\!\! s{\bm\alpha _{mn}}{\bm X_{mn}{ \bm w_{mn}}\!\!\left(\!\bm\alpha_{mn}\! \right)} \!\!+\!\!s{\pi_k}{\bm X_{mn}}{\bm w_{mn}}\!\!\left(\!\bm\alpha_{mn}\!\right)\!\!-\! \frac{\sigma}{2\mu_1}\!{{\left\| {s{\bm X_{mn}}{\left(\!{{\bm \pi}\!\!-\!\!{\bm \alpha_{mn}}}\right)} }\right\|}^2}}\!\!\right) \notag \\
=&\!\sum\limits_{n = 1}^N\!\!\left({\sum\limits_{k= 1}^{K}\!\!\left( {\!sl_{n}^*\!\left( {\!- {\alpha_{mn,k}}} \right) \!+\!\!s{\bm\alpha _{mn}} {\bm X_{mn}\bm w_{mn}\!\left( \bm\alpha_{mn}\right)}\!+\!\!s{l_{n}} ( \bm w_{mn}{{\left( \bm\alpha_{mn}\right)}^{\rm T}}{\bm x_{m,k}})} \right)\!\!+\!\frac{{\mu_2 (1\! -\! s)}}{2}s{{\left( {{\bm \pi}\!\!-\!{\bm\alpha_{mn}}} \right)}^2}}\right. \notag \\
&{\kern 10pt}\left.{ -\!\left\langle {{\bm w_{mn}}(\bm\alpha_{mn} ),\!s{\bm X_{mn}}\!\left( {\bm \pi}\!\!-\!\!{\bm\alpha _{mn}}  \right)} \right\rangle\!\!+\!\!\left\langle {{\bm w_{mn}}(\bm\alpha_{mn} ),\!s{\bm X_{mn}}\!\left(\bm \pi\!\!-\!\!\bm\alpha_{mn} \right) } \right\rangle\!\!-\!\!\frac{{\sigma}}{2\mu_1}{{\left\| s{\bm X_{mn}}\!{\left( {{\bm \pi}\!\!-\!\!{\bm\alpha_{mn}}} \right)}  \right\|}^2}} \! \right)  \notag \\
=&  sG\!\left(\bm\alpha_{mn}\right)\!\!+\!\! \sum\limits_{n = 1}^N \!{\left( {\frac{{\mu_2 (1 \!-\! s)}}{2}s{{\left( {{\bm \pi} \!\!-\!\! {\bm\alpha _{mn}}} \right)}^2}\!\!-\!\! \frac{{\sigma}}{2\mu_1}{{\left\| {{\bm X_{mn}}\left( {s\!\left( {{\bm \pi}\!\!-\!\!{\bm\alpha _{mn}}} \right)} \right)} \right\|}^2}}\! \right)}.
\end{align}\vspace{-0.1cm}

Here, under the assumption that $s=\frac{\mu_1\mu_2}{\mu_1\mu_2+\sigma} \in (0,1)$ and $\left\| \bm X_{mn}\right\|\leqslant 1$, it is easy to show that ${\left( {\frac{{\mu_2 (1 \!-\! s)}}{2}s{{\left( {{\bm \pi} \!\!-\!\! {\bm\alpha _{mn}}} \right)}^2}\!\!-\!\! \frac{{\sigma}}{2\mu_1}{{\left\| {{\bm X_{mn}}\left( {s\!\left( {{\bm \pi}\!\!-\!\!{\bm\alpha _{mn}}} \right)} \right)} \right\|}^2}}\! \right)}\geqslant0$. This completes the proof.
\vspace{-0.1cm}
\subsection{Proof of Theorem 2}
\vspace{-0.1cm}
Rewrite Lemma $3$ as:
\begin{equation}\label{eq:a53}
\begin{aligned}
\mathbb{E}\!\left[ {D\left( {{\bm\alpha_m^{(h)}}} \right) \!-\! D\left( {{\bm\alpha_m ^{(h+1)}}} \right)} \right] &= D\left( {{\bm\alpha_m ^{(h)}}} \right) \!-\! D\left( {{\bm\alpha_m^*}} \right) \!+\! E\!\left[ {D\left( {{\bm\alpha_m^{\rm{*}}}} \right) \!-\! D\left( {{\bm\alpha_m^{(h+1)}}} \right)} \right] \\
&\geqslant s\eta\! \left( {1 \!-\! \Theta } \right)\!G\left( {{\bm\alpha_m ^{(h)}}} \right) \\
&\geqslant s\eta\! \left( {1 \!-\! \Theta } \right)\left( {D\left( {{\bm\alpha_m^{(h)}}} \right) \!-\! D\left( {{\bm\alpha_m^{\rm{*}}}} \right)} \right),
\end{aligned}\vspace{-0.1cm}
\end{equation}
where the last inequality follows from the fact that $G\left( {{\bm\alpha_m^{(h)}}}\right) \!\geqslant\! \left({D\left( {{\bm\alpha_m^{(h)}}} \right) \!-\! D\left( {\bm \alpha_m^*} \right)} \right)$. Thus, (\ref{eq:a53}) can be rewritten as:
\vspace{-0.2cm}
\begin{equation}\label{eq:a54}
\begin{aligned}
\mathbb{E}\!\left[ {\left( {D\left( {{\bm\alpha_m^{h{\rm{ + }}1)}}} \right) \!-\! D\left( {{\bm\alpha_m^{\rm{*}}}} \right)} \right)} \right] \!\leqslant\! \left( {1 \!-\! s\eta \left( {1 \!-\! \Theta } \right)} \right)\!\left( {D\left( {{\bm\alpha_m^{(h)}}} \right) \!-\! D\left( {{\bm\alpha_m ^{\rm{*}}}} \right)} \right).
\end{aligned}\vspace{-0.1cm}
\end{equation}
Applying this inequality recursively for $h$ times and taking expectations from both sides, we have:
\vspace{-0.2cm}
\begin{equation}\label{eq:a54}
\begin{aligned}
\mathbb{E}\!\left[ {\left( {D\left( {{\bm\alpha_m^{(h+1)}}} \right) \!-\! D\left( {{\bm\alpha_m^{\rm{*}}}} \right)} \right)} \right] \!\leqslant\! {\left( {1 \!-\! s\eta\! \left( {1 \!-\! \Theta } \right)} \right)^{h + 1}}\!\left( {D\left( {{\bm\alpha_m^{(0)}}} \right) \!-\! D\left( {{\bm\alpha_m^{\rm{*}}}} \right)} \right).
\end{aligned}\vspace{-0.1cm}
\end{equation}
This completes the proof.

\renewcommand{\baselinestretch}{1.49}
\bibliographystyle{IEEEbib}
\bibliography{TCMultitask}

\end{document}